\documentclass[11pt,a4paper,dvipsnames]{article}
\usepackage{jheppub}
\usepackage{amsmath,amsfonts,relsize}
\usepackage[nameinlink,capitalize]{cleveref} 
\usepackage{graphicx}
\usepackage{dcolumn}
\usepackage[normalem]{ulem}\usepackage[mathlines]{lineno}

\usepackage{tabularx}
\usepackage{bm,color,xcolor}
\usepackage[font=small]{caption}
\usepackage{subcaption}
\usepackage{tablefootnote}

\usepackage{floatrow}
\usepackage{comment}

\usepackage{feynmp} 
\usepackage[compat=1.1.0]{tikz-feynhand}
\usepackage[compat=1.1.0]{tikz-feynman}
\usepackage[rightcaption]{sidecap}
\sidecaptionvpos{figure}{c}
 
\newfloatcommand{capbtabbox}{table}[][\FBwidth]

\usepackage{enumitem}

\usepackage{xspace}

\usepackage{slashed,physics} 
\usepackage{float}
\usepackage{multirow}
\usepackage{mathtools}
\usepackage{appendix}
\usepackage[colorlinks=true
,urlcolor=blue
,anchorcolor=blue
,citecolor=blue 
,filecolor=blue
,linkcolor=blue
,menucolor=blue
,linktocpage=true
,pdfproducer=medialab
,pdfa=true
]{hyperref}
\usepackage{bm,color,color,soul}
\definecolor{UMNRed}{RGB}{144, 0, 33}
\definecolor{UMNGold}{RGB}{255 204 51}

\newcommand{\D}{{\rm d}}
\newcommand{\es}[2] {\begin{equation} \label{#1} \begin{split} #2 \end{split} \end{equation}}
\newcommand{\mpl}{M_{\rm Pl}}

\newcommand{\beq}{\begin{equation}}
\newcommand{\eeq}{\end{equation}}
\newcommand{\bea}{\begin{eqnarray}}
\newcommand{\eea}{\end{eqnarray}}

\usepackage[arrows]{ezedits} 

\defineEditQuick{ZL}{UMNRed}
\newcommand{\SK}[1]{\textcolor{blue} {SK: #1}}

\title{Heavy QCD Axions at High-Energy Muon Colliders
}

\author[a]{Ravneet Bedi,}
\author[a]{Tony Gherghetta,}
\author[b,c]{Soubhik Kumar,}
\author[a]{Peiran Li,}
\author[a]{Zhen Liu,}

\affiliation[a]{School of Physics and Astronomy, University of Minnesota, Minneapolis, Minnesota 55455, USA}
\affiliation[b]{Institute of Cosmology, Department of Physics and Astronomy, Tufts University, Medford, Massachusetts 02155, USA}
\affiliation[c]{Center for Cosmology and Particle Physics, Department of Physics, New York University, New York, New York 10003, USA}

\emailAdd{bedi0019@umn.edu}
\emailAdd{tgher@umn.edu}
\emailAdd{soubhik.kumar@tufts.edu}
\emailAdd{li001800@umn.edu}
\emailAdd{zliuphys@umn.edu}

\preprint{\small UMN-TH-4507/25}

\abstract{
We study the physics potential of heavy QCD axions at high-energy muon colliders. Unlike typical axion-like particles, heavy QCD axions solve the strong CP problem with phenomenology driven by the anomalous gluon ($aG\widetilde G$) couplings.
Several ultraviolet scenarios are presented in which QCD axions with TeV-scale masses and decay constants arise consistently with a solution to both the strong CP problem and the axion quality problem.
We perform a detailed collider analysis for both a 3 and 10~TeV muon collider, focusing on hadronic axion decays that gives rise to a dijet-resonance signature. Our projections for the axion discovery reach in the multi-TeV mass range demonstrate that a muon collider can significantly extend sensitivity to heavy QCD axions compared to existing experiments.
}

\begin{document}
\maketitle

\section{Introduction}
The absence of observable CP violation in the strong sector of the Standard Model (SM) poses a long-standing theoretical challenge known as the strong CP problem. Although the QCD Lagrangian admits a CP-violating $\bar{\theta}$-term, measurements of the neutron electric dipole moment constrain $\bar\theta\lesssim10^{-10}$~\cite{Abel:2020pzs,Pendlebury:2015lrz,Baker:2013iya,Baker:2006ts}, requiring extreme fine-tuning without a known symmetry explanation. The most compelling solution to this puzzle is the Peccei–Quinn (PQ) mechanism~\cite{Peccei:1977hh,Peccei:1977ur}, which dynamically relaxes $
\bar\theta$ to zero through the introduction of a spontaneously broken global $U(1)_\text{PQ}$ symmetry. The associated pseudo-Nambu-Goldstone boson~\cite{Weinberg:1977ma,Wilczek:1977pj}, the axion, acquires a small mass via QCD non-perturbative effects (see for example~\cite{diCortona:2015ldu}).
This QCD axion, as originally formulated, is a light pseudo-Nambu-Goldstone boson with a mass and couplings inversely proportional to the $U(1)_\text{PQ}$ breaking scale $f_a$. 
Given the minimality of the axion solution to the strong CP problem and the fact that it can explain the observed DM abundance~\cite{Preskill:1982cy,Dine:1982ah,Abbott:1982af}, it represents one of the prime targets of physics beyond the SM.
Consequently, a large variety of astrophysical, cosmological, and laboratory probes continue to search for the QCD axion~\cite{ParticleDataGroup:2024cfk,Baryakhtar:2025jwh}. While there exist stronger bounds for certain ranges of QCD axion mass, various astrophysical constraints impose a robust lower limit $f_a \gtrsim 10^8$~GeV~\cite{Raffelt:2006cw, ParticleDataGroup:2024cfk}.
Since the QCD axion mass $m_a^{\rm QCD}$ is determined uniquely by $f_a$, the above constraint implies $m_a^{\rm QCD} \lesssim 60$~meV. 

An undesirable feature of minimal models, where the QCD axion arises as an elementary pseudo-Nambu-Goldstone boson from $U(1)_{\rm PQ}$ breaking, is the so-called ``axion quality problem"~\cite{Ghigna:1992iv, Holman:1992us, Kamionkowski:1992mf, Barr:1992qq}.
This problem stems from the fact that the QCD axion potential is rather delicate,
where even Planck scale-suppressed higher-dimensional operators can significantly modify the QCD axion potential, destabilizing the axion away from the desired $\langle a\rangle/f_a=\bar{\theta}$ minimum.
Since this would reintroduce the strong CP problem, a robust solution requires a mechanism that either suppresses these higher-dimensional operators or 
modifies the QCD axion potential while maintaining $\langle a\rangle/f_a=\bar{\theta}$ at its minimum.
A solution of the first type is found in scenarios with a large discrete symmetry, composite axions, and extra-dimensional/string theory axions; for a review see~\cite{DiLuzio:2020wdo}.
In this work, we pursue the second type of solution, which allow us to compute the QCD axion potential and its couplings to the SM reliably within the regime of 
conventional field theory, and do not require the details of string compactifications.
By modifying the QCD axion potential, for a given value of $f_a$, the QCD axion mass can be larger, therefore giving rise to so-called ``heavy QCD axions".
An axion mass larger than ${\cal O}(100)$~MeV implies that astrophysical processes, such as those occurring in supernovae, cannot produce such heavy axions and consequently much smaller values of $f_a$ are observationally allowed.
In particular, the regime $m_a\sim$~100~MeV--10~TeV with $f_a \sim$~100~GeV--PeV is ripe for exploration; this parameter space can be probed effectively at both the LHC~\cite{Jaeckel:2012yz,ATLAS:2022abz,ATLAS:2024bjr,CMS:2024yhz} and beam-dump environments~\cite{CHARM:1985anb,Dolan:2017osp,Dobrich:2019dxc,NA64:2020qwq,Capozzi:2023ffu, Brdar:2020dpr, Kelly:2020dda, Blinov:2021say, Co:2022bqq}.
At the same time, in some of this parameter space, including $m_a \gtrsim$~TeV and $f_a\gtrsim$~TeV, the axion potential is modified 
to the extent that even dimension-five Planck-suppressed operators do not destabilize the axion away from the desired minimum.
Thus, in this regime, both the strong CP problem and the axion quality problem can be solved simultaneously.

While several previous studies~\cite{Jaeckel:2012yz,Mariotti:2017vtv,Hook:2019qoh, Knapen:2021elo, Ghebretinsaea:2022djg} have investigated the discovery prospect of heavy QCD axions at the LHC, for $m_a \gtrsim$~TeV, the LHC sensitivity rapidly degrades, as expected.
Given the importance of this parameter space, in this work, we instead investigate the discovery potential of a future 10 TeV muon collider (MuC) for heavy QCD axions. 
As theory benchmarks, we examine several classes of ultraviolet(UV) models that give rise to heavy QCD axions with $m_a$ and $f_a$ in the desired range. 
To understand quantitatively the sensitivity of a MuC, we then carry out detailed MuC simulation of the signal and the associated SM background processes.
We note that axion-like particle (ALP) phenomenology at muon colliders has been studied in the literature~\cite{Casarsa:2021rud,Bao:2022onq,Han:2022mzp,Inan:2022rcr,Chigusa:2023rrz,Chigusa:2025otr,Bose:2022obr} primarily through their couplings to electroweak gauge bosons and heavy quarks.
On the other hand, in our work, the focus is exclusively on heavy QCD axions that can solve the strong CP problem. Consequently, the defining axion-gluon coupling plays a crucial role in determining the resulting phenomenology.
While the heavy QCD axion production is dominated by electroweak couplings, similar to ALPs, the primary decay channel is into a pair of gluons.
We demonstrate that the clean environment in a muon collider provides a uniquely powerful probe of heavy QCD axions leading to a di-jet final state, with sensitivity well beyond the reach of current
facilities, including the high-luminosity LHC (HL-LHC). 
Thus, high-mass dijet searches at a MuC, while interesting in themselves, are intimately connected with potential solutions to the strong CP problem and the axion quality problem at terrestrial experiments.

To further highlight the potential reach and unique capability of the MuC,
we briefly summarize some of the existing collider analyses for ALPs, which exhibit similar couplings and decay channels to those of the heavy QCD axion but do not necessarily address the strong CP problem. 
Lepton colliders have been used to search for ALPs, with sensitivity reaching masses up to $\sim$100~GeV~\cite{Jaeckel:2015jla,OPAL:2000puu,Pustyntsev:2023rns,BESIII:2022rzz,BESIII:2024hdv,Belle-II:2020jti}. 
At the LHC, heavier ALPs with gluonic couplings have been explored mainly through di-photon resonance searches~\cite{Jaeckel:2012yz,Mariotti:2017vtv, ATLAS:2017ayi, Bauer:2018uxu,Gershtein:2020mwi, ATLAS:2022abz, ATLAS:2024bjr,CMS:2024yhz}, extending the mass reach to the TeV scale, and as well tens of GeV scale through long-lived particles signatures~\cite{Hook:2019qoh}. However, searching for heavy QCD axions is difficult due to the large QCD background at a hadron collider. 
Future $e^+e^-$ colliders such as the CEPC and FCC-ee will provide improved sensitivity to ALPs or QCD axions near the electroweak regime~\cite{Bernardi:2022hny,CEPCStudyGroup:2018ghi,CEPCPhysicsStudyGroup:2022uwl,ILCInternationalDevelopmentTeam:2022izu,RebelloTeles:2023uig,Bao:2025tqs}.  
Nevertheless, probing axions with masses in the multi-TeV range—particularly those originating from models that modify QCD at UV scales and still solve the strong CP problem—requires next-generation colliders with substantially higher center-of-mass energies. For example, at the FCC-hh, heavy ALPs have also been studied~\cite{Bauer:2018uxu,FCC:2025lpp}, covering a wide mass range from the GeV to $\mathcal{O}(10)~$TeV scale.

The paper is organized as follows. We first provide a detailed study of heavy QCD axion phenomenology at a muon collider in \autoref{sec:muc_probe}, including signal production, background simulations, and projected sensitivities. In \autoref{sec:axion_model}, several UV models that predict a heavy QCD axion around the multi-TeV scale are presented. Finally, in \autoref{sec:conclusions}, we summarize our findings and discuss future prospects for probing axions at muon colliders. Additional details of the axion production calculation are given in \autoref{sec:calculation_VBS}.

\section{Probing Heavy QCD Axions at a Muon Collider}
\label{sec:muc_probe}
In this section, we present a general study of heavy QCD axion phenomenology at high-energy muon colliders.
A future high-energy muon collider combines a high center-of-mass energy up to 10~TeV and a low-background environment due to its comparatively low rate for QCD final states. In particular, we focus on the 10~TeV (3~TeV) MuC scenario with 10~$\text{ab}^{-1}$ (1~$\text{ab}^{-1}$) integrated luminosity. 
As will be shown in \autoref{sec:axion_model}, several heavy QCD axion models, can be uniquely examined at the MuC that covers a large fraction of the $f_a$-$m_a$ parameter space. 

Heavy QCD axions commonly interact with the Standard Model gauge bosons via dimension-five anomaly-induced couplings. 
The effective Lagrangian that governs these interactions is given by
\begin{align}
    \mathcal{L}\supset c_3\frac{\alpha_s}{8\pi f_a}aG\widetilde{G}+c_2\frac{\alpha_2}{8\pi f_a}aW\widetilde{W}+c_1\frac{\alpha_1}{8\pi f_a}aB\widetilde{B},
    \label{eq:axioneffL}
\end{align}
where $a$ denotes the axion field, $f_a$ is the axion decay constant, $c_i$ are the anomaly coefficients and $\alpha_{1,2,s}$ are the SM gauge couplings associated with the $U(1)_Y$, $SU(2)_L$, $SU(3)_c$ gauge groups, respectively. The dual field strengths $\widetilde{G}, \widetilde{W}, \widetilde{B}$ encode the CP-odd structure of these interactions.
Due to the effective interactions in \eqref{eq:axioneffL}, the production of heavy QCD axions at a muon collider is dominated by electroweak processes, while QCD processes dominate their decay. This provides a distinct difference between previous studies of heavy ALP phenomenology at colliders.
Since the effective Lagrangian is valid below the scale $f_a$, we can first study the heavy QCD axion phenomenology with the generic interactions terms in \eqref{eq:axioneffL} and then match to specific UV models in \autoref{sec:axion_model}.

\subsection{Heavy Axion Production and Decays}
\label{sec:production}
Heavy QCD axions that couple to electroweak bosons can be produced at a MuC and then decay into a pair of gluons. In the following we discuss the prominent production and decay modes.

\subsubsection{Production Modes} 
\begin{figure}
    \centering
    \includegraphics[width=0.2\textwidth]{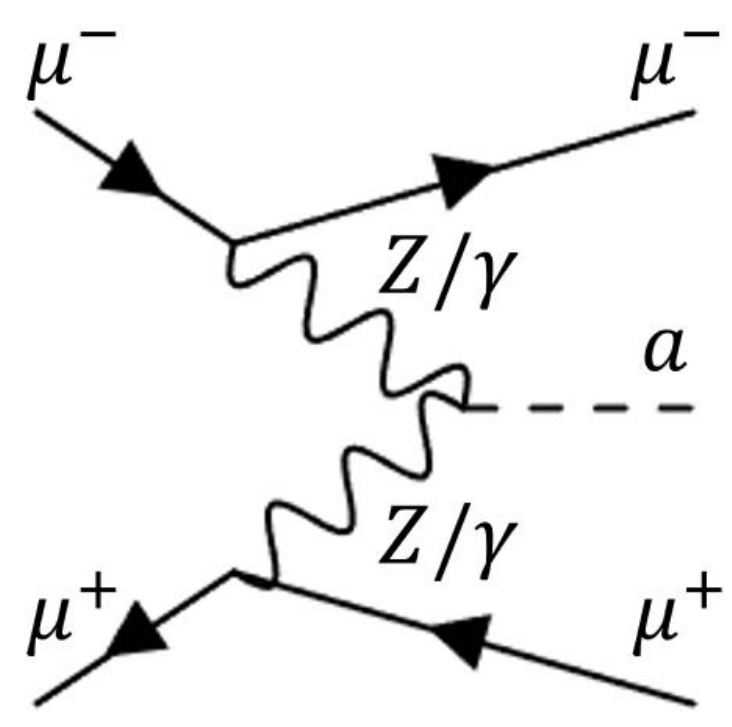}~~~~~~
    \includegraphics[width=0.2\textwidth]{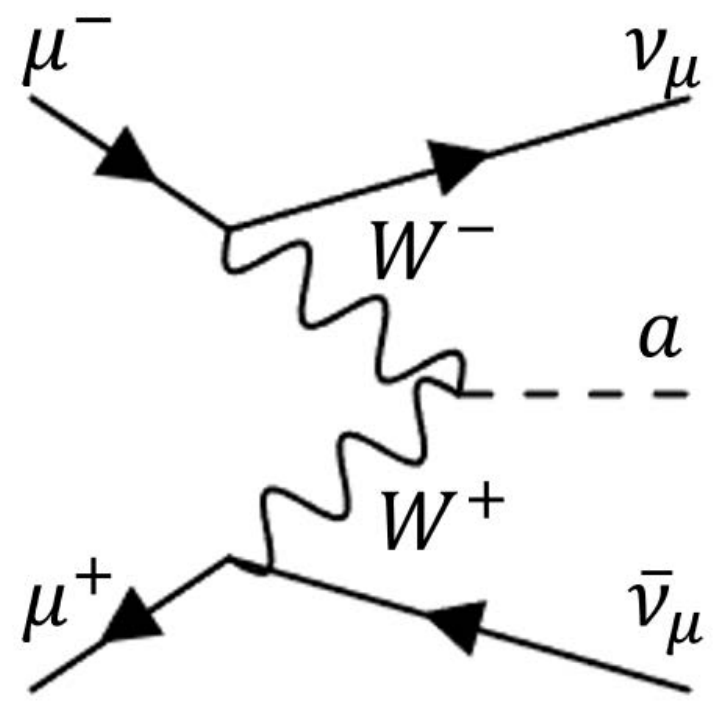}~~~~~
    \includegraphics[width=0.27\textwidth]{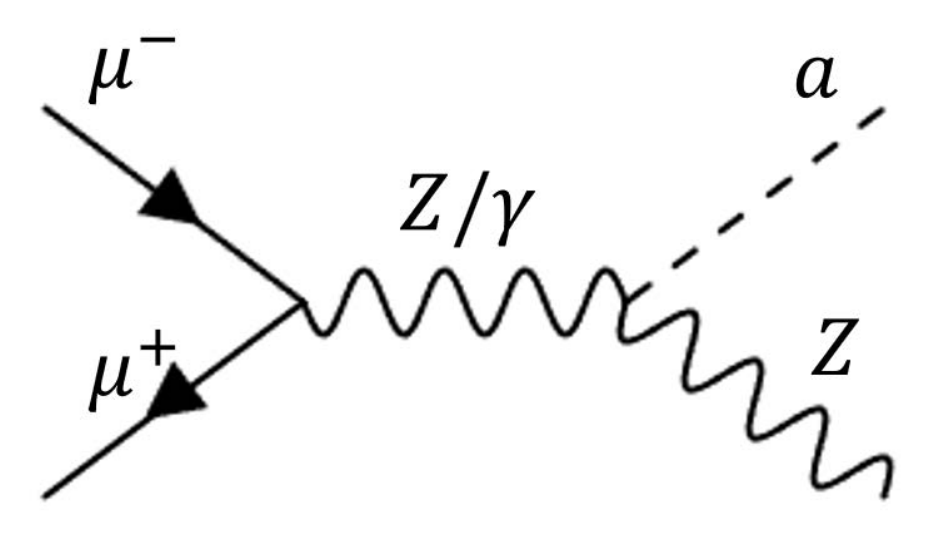}\\
    \vspace{3ex}
    \includegraphics[width=0.26\textwidth]{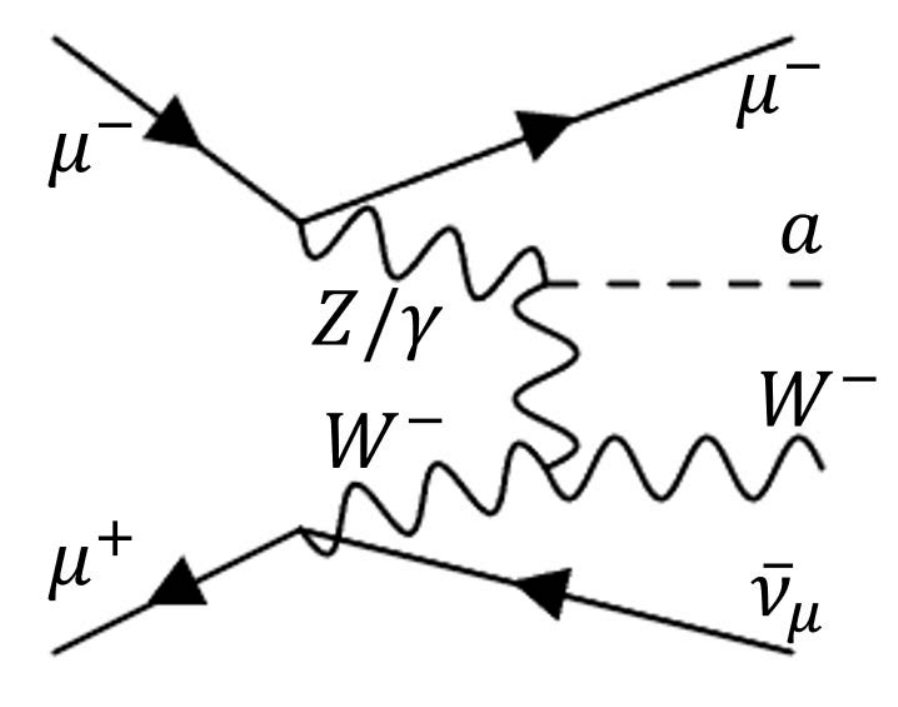}~~~~~
    \includegraphics[width=0.26\textwidth]{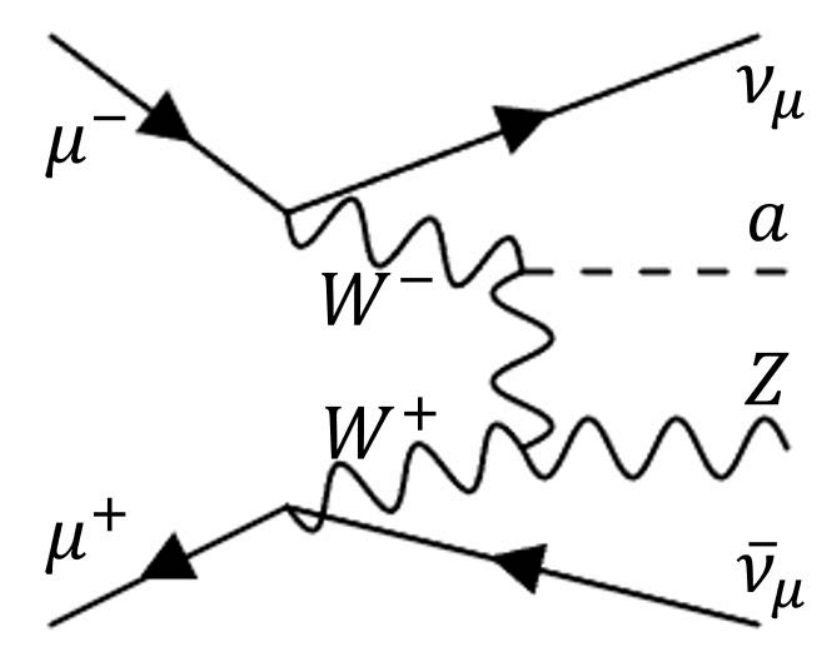}
    \caption{Selected representative Feynman diagrams corresponding to the axion production channels due to the dominant VBF production (top left), the subdominant VBF production (top middle), associated axion production (top right) and VBS production (bottom row).
    }
    \label{fig:tree_level_production_diagram}
\end{figure}

We include both vector-boson-fusion (VBF) production and associated axion production with a vector boson. Additionally, we include vector-boson-scattering (VBS) channels which also contribute significantly at a 10 TeV MuC. 
Overall, as depicted in \autoref{fig:tree_level_production_diagram}, there are five main axion production modes: (1) $\mu^+ \mu^- \to a+ \mu^+ \mu^-$ (dominant VBF production), (2) $\mu^+ \mu^- \to a+ \nu_\mu \bar{\nu}_\mu $ (subdominant VBF production), (3) $\mu^+ \mu^- \to a+ Z$ (associated axion production), (4) $\mu^+ \mu^- \to a + \mu^\pm \nu_\mu W^\mp$ (VBS production), and (5) $\mu^+ \mu^- \to a + \nu_\mu \bar{\nu}_\mu Z $ (VBS production).

\begin{figure}
    \centering
    \includegraphics[width=0.9\textwidth]{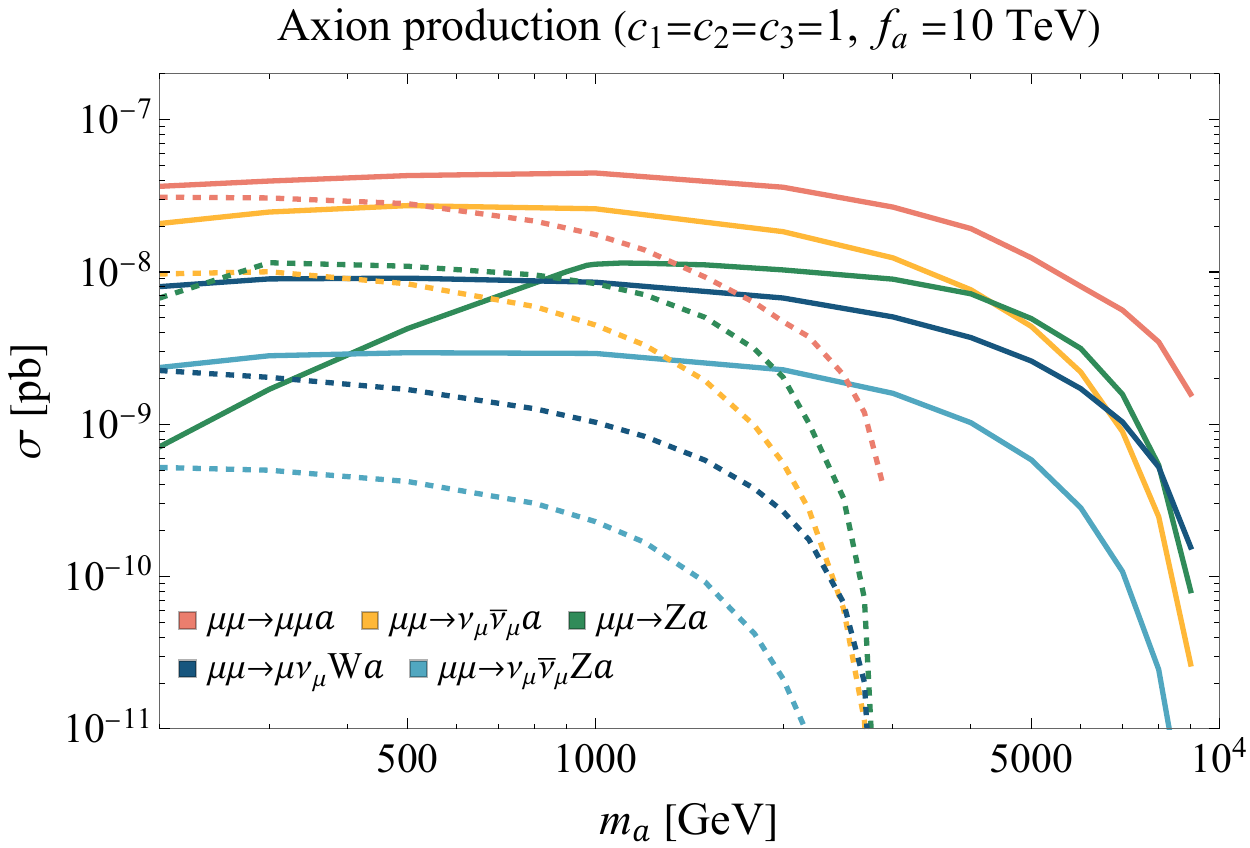}
    \caption{The cross section $\sigma$ of the main axion production channels as a function of the axion mass $m_a$ with central acceptance cuts on the gluon jet $|\eta(g)|<2.5, p_T(g)>10~\text{GeV},\Delta R(gg)>0.4$ at a 3 (dashed curves) and 10~TeV (solid curves) muon collider.
    }
    \label{fig:xs_signal}
\end{figure}

The cross sections of these five production modes as a function of the axion mass are shown in \autoref{fig:xs_signal}, where no special hierarchies between the Wilson coefficients, $c_i$ have been assumed.
The production rate at muon colliders is determined by the electroweak couplings $c_1$ and $c_2$. We can see that at both the 3 TeV and 10 TeV MuC, the VBF processes dominate the heavy axion production. 

The cross section can be obtained by either a fixed-order calculation with proper IR regulators or an electroweak parton distribution function (PDF) calculation, each with its own advantages and limitations. We have performed both calculations for all the VBF and VBS processes. In the main text, for simplicity, we focus on the electroweak PDF explanation as it allows for partonic scattering discussions which are lower-order processes. 
Note that the fixed-order calculation is unambiguous with all interferences between diagrams taken into account consistently, once the IR is regulated properly, in particular for those processes involving internal photons. It serves as a cross-check and provides guidance on the appropriate matching scales for the underlying physical processes. 
For more details of the cross-check between the different methods, see \autoref{sec:calculation_VBS} for both VBF and VBS processes. 
In \autoref{fig:xs_signal}, the dominant VBF and the VBS production channels are evaluated using the electroweak PDF treatment to simplify the matching and regulator scale discussion. The subdominant VBF production does not have collinear photon splittings, since it is regulated by the $W$ boson mass. Therefore, both $\mu^+\mu^- \to \nu_\mu \bar{\nu}_\mu a$ and $\mu^+\mu^- \to Z a$ processes are calculated at fixed order.
\vspace{3mm}

\noindent
Next, we will explain the individual production channels in more detail:
\vspace{-3mm}
{\flushleft (1) \underline{$\mu^+\mu^-\to \mu^+\mu^-a$}}

\noindent
This channel is dominated by the photon PDF with the sub-processes:
$\gamma^*\gamma^*/\gamma^* Z^*/ Z^*Z^* \to a$.
We do not demand the tagging of forward muons, as the heavy dijet resonance itself is powerful enough to distinguish the signal from background in the clean muon collider environment. As shown in the detailed comparison calculation in \autoref{fig:fixed_order_xs} of \autoref{sec:calculation_VBS}, the fixed-order calculation matches the electroweak PDF calculation with a GeV matching scale. 

{\flushleft (2) \underline{$\mu^+\mu^- \to \nu_\mu \bar{\nu}_\mu a$}}

\noindent
This channel is dominated by $WW$-fusion: $W^* W^*\to a$.
The $W$ boson PDF is much smaller than the photon PDF, which can be understood by the leading logarithmic approximation spectrum~\cite{Lindfors:1986sh,Han:2020uid,Han:2021kes,Ruiz:2021tdt}
\begin{align}
    f_{\gamma / \mu}(x)&=\frac{\alpha}{2\pi}\frac{1+(1-x)^2}{x} \ln{\left(\frac{Q^2}{m_\mu^2}\right)},\\
    f_{W_T / \mu}(x)&=\frac{\alpha_2}{4\pi}\frac{1+(1-x)^2}{x} \ln{\left(\frac{Q^2}{m_W^2}\right)},\\
    f_{W_L / \mu}(x)&=\frac{\alpha_2}{2\pi}\frac{1-x}{x}.
\end{align}
However, the effective coupling term $aW\widetilde{W}$ contains $\alpha_2$ which is roughly four times larger than the electromagnetic coupling $\alpha$. Despite the sizable logarithmic enhancement in the photon PDF, due to the difference in coupling strength, the $WW$-fusion rate at muon colliders is comparable to the photon fusion rate within an order of magnitude.
Note that naively, the longitudinal $W/Z$ boson fusion may significantly contribute to the axion production rate since its polarization vector scales as $\frac{p^\mu}{m_{W/Z}}$. However, the antisymmetric tensor structure in the axion coupling and the forward kinematics of the $W/Z$ bosons from the muon PDF forbid such longitudinal enhancements. The partonic helicity amplitude is proportional to $p_1^\alpha p_2^\beta \epsilon_L^\mu(p_1) \epsilon_{T/L}^\nu(p_2)\epsilon_{\alpha\beta\mu\nu}$, which vanishes when the $W/Z$ bosons are treated as on-shell and at least one of them is longitudinally polarized. For example, 
\begin{align*}
    p_1^\alpha p_2^\beta \epsilon_L^\mu(p_1) \epsilon_{T(L)}^\nu(p_2)\epsilon_{\alpha\beta\mu\nu}&=\det \begin{pmatrix}
    p_1^0 &p_2^0 &\frac{\sqrt{(p_1^0)^2-m_W^2}}{m_W} & ~~~0 \,\left(\frac{\sqrt{(p_2^0)^2-m_W^2}}{m_W}\right)  \\
    0 & 0 & 0 & \frac{1}{\sqrt{2}}~(0)  \\
    0 & 0 & 0 & \frac{i}{\sqrt{2}}~(0)  \\
    \sqrt{(p_1^0)^2-m_W^2} &-\sqrt{(p_2^0)^2-m_W^2} &\frac{p_1^0}{m_W} & 0~\left(\frac{p_2^0}{m_W}\right)  \\
    \end{pmatrix}\\
    &= 0\,,
\end{align*}
where $p_1,p_2$ are the momenta of the $W/Z$ boson radiated from the muon and the parenthesis in the last column of the matrix denotes the case when the second gauge boson is also longitudinally polarized. Thus, from the perspective of the PDF, only the transverse modes contribute to the VBF production processes.

{\flushleft (3) \underline{$\mu^+\mu^- \to Z a$}}

\noindent
This is the $Za$ associated production channel which suffers from $s$-channel suppression. As shown in \autoref{fig:xs_signal}, the axion associated production has a kink around 1~TeV because we require a sizable angular separation $\Delta R$ between the two jets. Without the $\Delta R$ requirement, the associated production would be a straight line in the low mass regime. For axion masses above a TeV, the two jets from its decay have enough momentm to be well separated. Moreover, when imposing the requirement to reconstruct both a $Z$ boson and a high-mass dijet resonance, the background is significantly reduced compared to a VBF production search. One can also include $\gamma a$ associated production, but it also leads to significantly more SM background with a hard photon in the final state and hence is not included in our current study. 

{\flushleft (4) \underline{$\mu^+ \mu^- \to \mu^\pm \nu_\mu W^\mp a$}}

\noindent
This is a higher-order VBS process compared to VBF production, driven by the subprocess $\gamma^* W^* \to a + W$. At first glance, the amplitude of this process may appear to scale as $E^4$, due to the involvement of three longitudinal $W/Z$ bosons and momentum-dependent couplings.
However, as discussed in the second channel (2), the axion does not couple efficiently to longitudinal modes. This implies that the on-shell gauge boson connected to the axion in the diagram must be transverse, reducing the energy scaling of the amplitude to at most $E^3$. Furthermore, when all relevant diagrams are summed, a significant cancellation\footnote{
While FeynRules correctly generates the Feynman rules, we found that the Universal FeynRules Output (UFO) file produced by FeynRules contains an incorrect sign in the contact interaction term $aW^+W^-Z/\gamma$. This sign error affects the expected cancellation among diagrams. To obtain correct results in the MadGraph simulation, we manually corrected the sign of the relevant coupling in the UFO file.} 
occurs, and the total amplitude scales only linearly with energy $E$. This cancellation of the $E^3$ scaling down to $E$ scaling is explicit in the unitary gauge. On the other hand, the result can be independently cross-checked using the Goldstone equivalence theorem. In the unitary gauge calculation, the leading two diagrams that scale as $E^3$ involve two longitudinal gauge bosons, the axion, and one transverse gauge boson. The corresponding Goldstone diagram for this process, however, exhibits only linear scaling with $E$.  
The diagram using Goldstone equivalence arises from the three-point $a$-$W_T$-$W_T$ coupling and the gauge kinetic term of the Goldstone bosons with a transverse $W_T$ boson. 
This behavior is similar to the well-known case of $W_L W_L \to W_L W_L$ scattering in the SM, where the leading energy growth cancels out among diagrams due to gauge symmetry.
Nevertheless, at a 10 TeV muon collider, this channel has a cross section significantly larger than that of the $Za$ associated production for axion masses below 1~TeV (see \autoref{fig:xs_signal}). 

{\flushleft (5) \underline{$\mu^+ \mu^- \to \nu_\mu \bar{\nu}_\mu Z a$}} 

\noindent
This channel exhibits similar behavior to (4), but with the photon PDF replaced by the $W$-boson PDF, resulting in a generally smaller cross section than that of the fourth channel.

\subsubsection{Decay Modes}

Next, we briefly discuss the heavy axion decays considered in this work. The decay widths of the main axion decay channels are: 
\begin{align}
    &\Gamma_{a\to gg}=8\times\frac{1}{2}\times\alpha_s^2 c_3^2\,\frac{m_a^3}{128\pi^3 f_a^2},\\
    &\Gamma_{a\to \gamma\gamma}=\frac{1}{2}\times\alpha^2 c_\gamma^2 \frac{m_a^3}{128\pi^3 f_a^2}~,\\
    &\Gamma_{a\to Z\gamma}=\alpha_{\gamma Z}^2 c_{\gamma Z}^2\,\frac{m_a^3}{128\pi^3 f_a^2}\left(1-\frac{m_Z^2}{m_a^2}\right)^{3/2}~,\\
    &\Gamma_{a\to ZZ}=\frac{1}{2}\times\alpha_Z^2 c_Z^2 \,\frac{m_a^3}{128\pi^3 f_a^2}\left(1-\frac{m_Z^2}{m_a^2}\right)^3~,\\
    &\Gamma_{a\to W^+ W^-}=\alpha_2^2 c_2^2 \,\frac{m_a^3}{128\pi^3 f_a^2}\left(1-\frac{m_W^2}{m_a^2}\right)^3~,
\end{align}
where we have defined
\begin{align}
    \alpha_{\gamma Z}=\alpha_2\tan\theta_W~,\qquad \alpha_Z=\alpha_1+\alpha_2~,
\end{align}
\begin{align}
    c_\gamma= c_1+c_2\,,\quad c_{\gamma Z}=2(c_2\cos^2\theta_W-c_1\sin^2\theta_W)\, \quad \text{and}\quad c_Z=c_2\cos^4\theta_W+c_1\sin^4\theta_W\,,
\end{align}
and $\theta_W$ is the weak mixing angle.
The main axion decay channel is the hadronic decay $a\to gg$, with a branching ratio $>0.95$ ($c_1=c_2=c_3=1$) due to the large QCD coupling $\alpha_s$ and color factors. Consequently, a di-jet final state is the main signature we are interested in.

\subsection{Simulation and Projections}
To provide projections at a MuC for heavy QCD axions, including various kinematics and resolution effects, we proceed with numerical simulation for the signal and backgrounds with event generation by {\tt MadGraph5\_aMC@NLO}~\cite{Alwall:2014hca}. The heavy axion model file in the Universal FeynRules Output (UFO)~\cite{Degrande:2011ua} is generated by FeynRules~\cite{Degrande:2011ua,Alloul:2013bka}. In this numerical study, the dominant VBF production, the VBS productions and one of the SM background $\mu^+ \mu^- \to \mu^+ \mu^- Wjj$ are obtained using effective vector boson approximation with the leading logarithmic resummation. The subdominant VBF production $\mu\mu\to\nu_\mu \bar{\nu}_\mu a$ does not receive large logarithmic enhancement and hence is simulated with the fixed-order treatment. 
The $Va$ associated production and all other background processes are also generated with a fixed-order calculation. The background considerations for these processes are also different. The $Va$ associated and VBS productions involve a visible $Z/W$ boson, whereas VBF productions do not; we treat these two sets of channels separately.

\begin{table}[htbp]
\centering
\resizebox{0.8\textwidth}{!}{%
\begin{tabular}{|c|c|c|}
\hline
BSM search & SM background process & Cross section\\
\hline
Axion VBF process& $\mu^+ \mu^- \to \mu^+ \mu^- jj$ & 0.47 pb \\
Axion VBF process& $\mu^+ \mu^- \to \nu_\mu \bar{\nu}_\mu  jj$ & $3.7\times 10^{-2}$ pb \\
Axion VBF process& $\mu^+ \mu^- \to \slashed{\gamma}/\slashed{Z}/\slashed{W}+jj$ & $3.0\times 10^{-3}$ pb \\
\hline
Axion associated+VBS process & $\mu^+ \mu^- \to \mu^+ \mu^- W+jj$ & $6.85\times 10^{-3}$ pb \\
Axion associated+VBS process & $\mu^+ \mu^- \to Z/W+jj$ & $2.69\times 10^{-3}$ pb \\
\hline
\end{tabular}%
}
\caption{Cross sections of the main background processes at a 10~TeV muon collider after imposing pre-selection via fixed-order calculation. For the $\mu\mu\to\mu\mu a/\nu_\mu\bar{\nu}_\mu a$ search, the first three backgrounds need to be considered, and the $\gamma/Z/W$ of the third background need to be out of the detector region $|\eta(\gamma/Z/W)|>2.5$ and hence missing (indicated with a slash notation). For the $\mu\mu\to 
Za/ \mu\nu Wa/\nu_\mu\bar{\nu}_\mu Z a$ search, we require the last two backgrounds including visible $Z/W$, where $|\eta(Z/W)|<2.5$\,.
}
\label{Table:bkg_xs_10TeV}
\end{table}

\begin{figure}[htbp]
    \centering
    \includegraphics[width=0.495\textwidth]{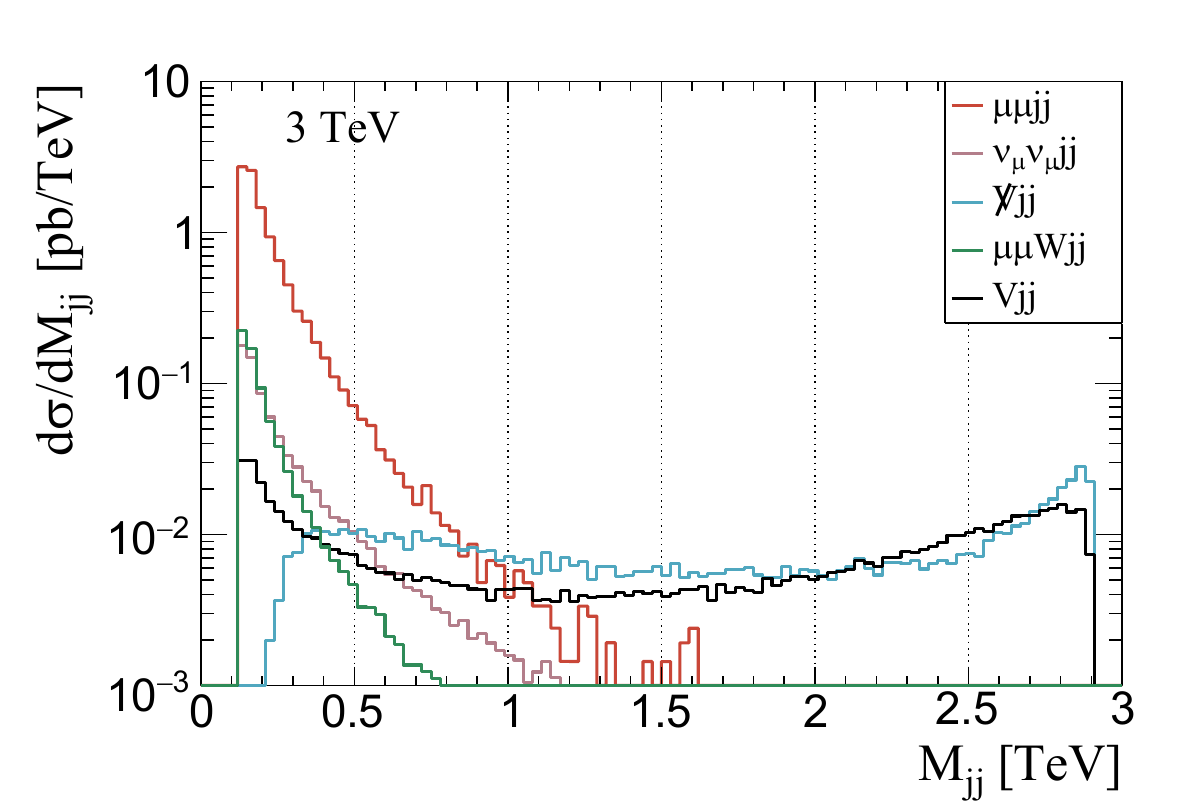}
    \includegraphics[width=0.495\textwidth]{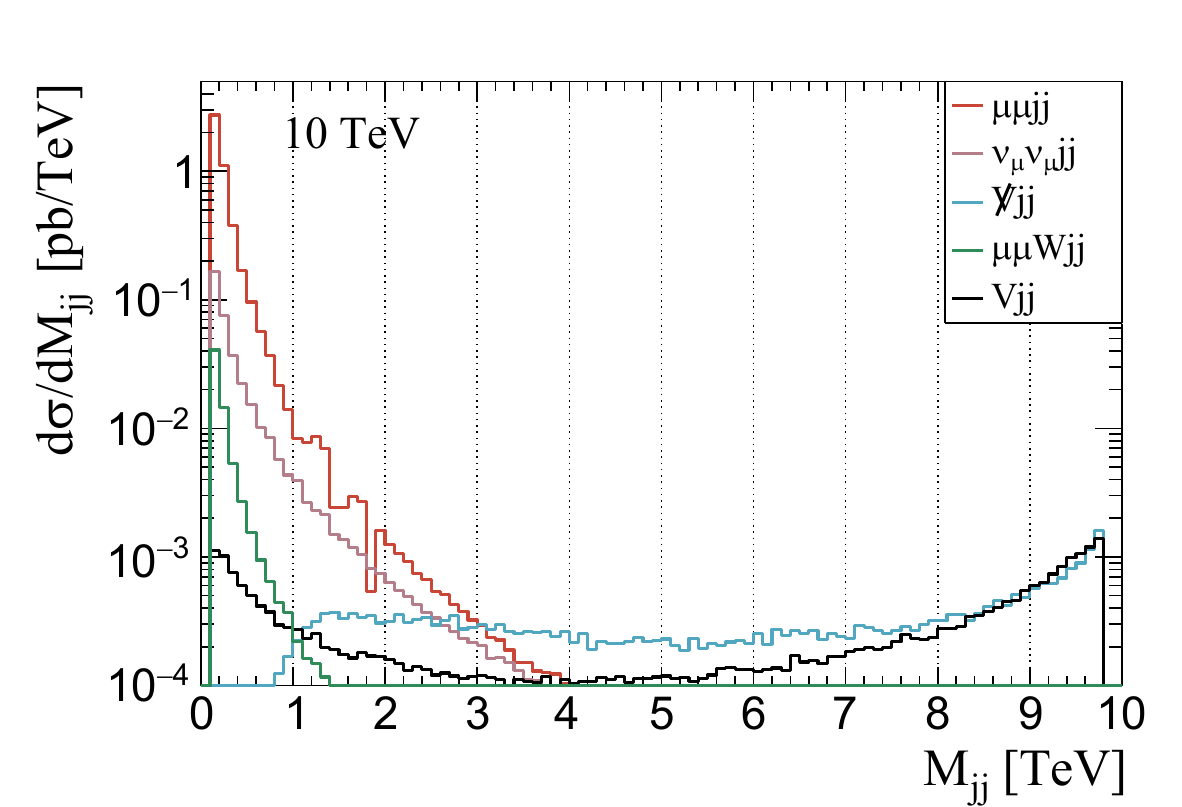}
    \caption{The di-jet invariant mass distribution of background processes after imposing pre-selection at a 3 and 10~TeV muon collider. The first three processes are the backgrounds for searching axion VBF production. The green and black curves refer to the backgrounds for associated production and VBS production of axion where $|\eta(Z/W)|<2.5$\,.}
    \label{fig:bkg_xs_10TeV}
\end{figure}

The background processes are tabulated in \autoref{Table:bkg_xs_10TeV}. Their di-jet invariant mass distributions are also shown in \autoref{fig:bkg_xs_10TeV}. The first two backgrounds are cumulated within low $\text{M}_{jj}$ since they are coming from vector boson scattering, dominated by low $Q^2$ intermediate vector bosons (or equivalently, low $x$ part of the electroweak PDF). The third and fifth backgrounds are mainly the Drell-Yan process with an initial radiation state or final radiation state consisting of $\gamma/Z/W$ boson. The fourth background can be viewed as a VBS process $\gamma^*\gamma^*\to Wjj$. 

The {\it pre-selection} for both signal and background is defined as
\begin{align*}
    130~\text{GeV}<\text{M}_{jj}<9800~\text{GeV},~\Delta R(\gamma j)>0.2,~\Delta R(jj)>0.4,~|\eta(j)|<2.5~\text{and}~p_T(j)>10~\text{GeV}
\end{align*}
where $\Delta R=\sqrt{(\Delta\phi)^2+(\Delta \eta)^2}$ and $j$ refers to jets originating from gluons or light quarks ($u,d,c,s$). After applying the pre-selection, we finally select events that have a di-jet invariant mass within 
$[0.95\,m_a,~1.05\,m_a]$ 
for a given benchmark value of $m_a$.  For the VBF production search, we require that only the dijet system falls in the central detector region within $|\eta|<2.5$, and veto other hard objects in the region. 
In contrast, for the VBS and associated production search, we require both the high-mass di-jet system and the reconstructible $Z/W$ boson to lie within the central region $|\eta|<2.5$. 
Both hadronic and (visible) leptonic decays of the $W/Z$ bosons is considered for the VBS and $Za$ signal processes, and we further assume an 80\% reconstruction efficiency of these gauge bosons. 
Here, we do not pursue further optimization as it would depend on the detailed detector performance, but this parton-level simulation with tolerance for uncertainty should provide us with a reasonable projection of the performance given the clean muon collider environment.

\begin{figure}[htbp]
    \centering
    \includegraphics[width=0.6\textwidth]{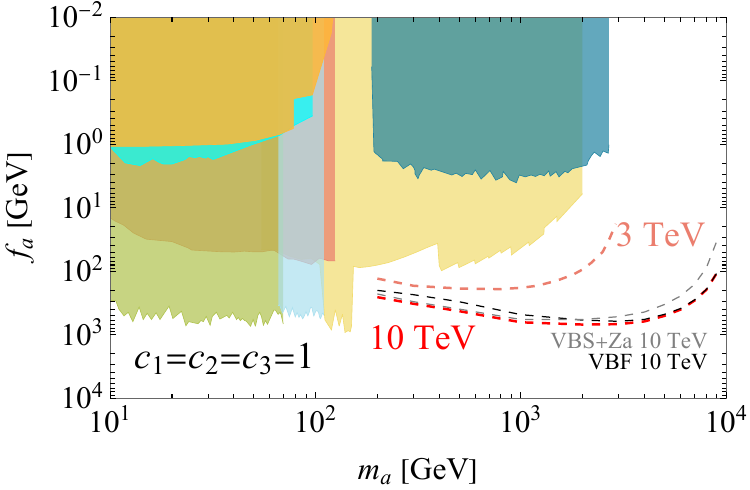}\\
    \includegraphics[width=0.6\textwidth]{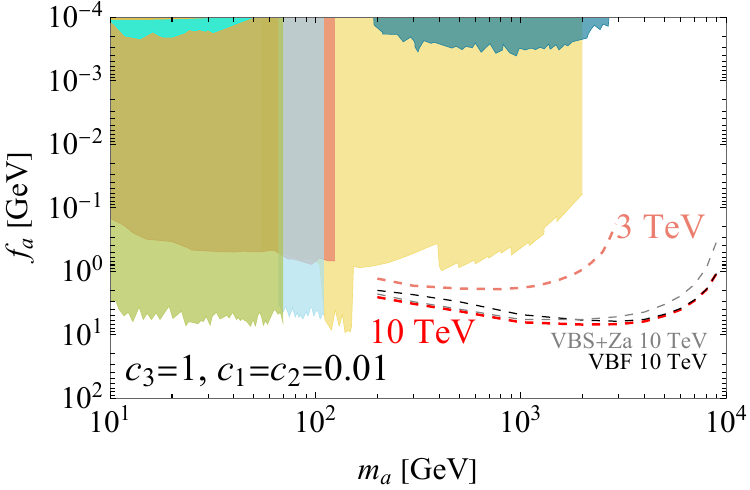}
    \caption{The projected sensitivity of a heavy QCD axion at a 3 and 10~TeV MuC shown as red-dashed lines for two different benchmark values of the couplings $c_i$. For comparison, the sensitivities of two separate channels at a 10 TeV MuC is shown as the gray and black lines. 
    The colored regions represent existing experimental constraints; for details, see the main text.
    }
    \label{fig:projection}
\end{figure}

The 95\% C.L. projected sensitivities of the axion decay constant $f_a$ as a function of $m_a$ at a 3 and 10 TeV MuC are shown in \autoref{fig:projection} as red-dashed lines. Additionally, the projected sensitivities of two individual channels (for a 10 TeV MuC) are also presented as gray and black dashed lines.
The production in general decreases as $m_a$ increases, but the dominant SM background resides in the dijet invariant mass $\sim 1~$TeV regime, where the electroweak process dominates. As a result, the sensitivity improves with increasing $m_a$ until the signal production starts to drop rapidly due to PDF suppression (see \autoref{fig:xs_signal}). Furthermore, the background for the VBF signal is significantly larger than that for the $Za$ and VBS channels, especially below a $1~\text{TeV}$ axion mass. Consequently, both the $Za$ associated production and VBS processes together provide the main contribution to the projected sensitivity below 1~TeV. 

The color-shaded regions are existing constraints. In the axion mass range from 10~GeV to 2~TeV, the main constraints (red, green, and yellow shaded regions) are due to the diphoton search (with gluon fusion) at the LHC~\cite{Jaeckel:2012yz,Mariotti:2017vtv,ATLAS:2022abz}. From 200~GeV to 2.7~TeV, the dark-blue limit is an exclusive diphoton search (with photon fusion) at the LHC~\cite{ATLAS:2017ayi,Bauer:2018uxu}. In the upper-left region, the orange and cyan shaded regions correspond to the $Z\to3\gamma$ search~\cite{Jaeckel:2015jla} at LEP and coherent photon fusion in Pb-Pb collisions~\cite{Knapen:2016moh,CMS:2018erd,ATLAS:2020hii}. Note that the constraints arising from electroweak production become significantly weaker as $c_1$ and $c_2$ decrease (which can happen for the extra-dimension model discussed in~\autoref{sec:axion_model}). In particular, we also reinterpret the most recent studies of the 95~GeV diphoton excess conducted by ATLAS and CMS~\cite{ATLAS:2024bjr,CMS:2024yhz} shown as the light-blue region. Although these constraints are extracted from current LHC results, the HL-LHC will improve the constraints on $f_a$ by a factor of $20^{1/4}\approx 2$ due to the improvement of statistics. In addition, a dijet search via gluon fusion at the HL-LHC~\cite{Ghebretinsaea:2022djg} could also yield powerful sensitivity, but this depends sensitively on the systematic uncertainties associated with jet measurements. 

In summary, muon colliders can probe heavy QCD axions with  $f_a\sim$~TeV, for ${\cal O}(1)$ Wilson coefficients  
(see upper panel of \autoref{fig:projection}), 
covering axion masses in the multi-TeV regime up to 10 TeV. 
Furthermore, the muon collider sensitivity depends linearly on the electroweak coupling coefficients, $c_1$ and $c_2$. Hence, when these coefficients are smaller, the sensitivity on the scale $f_a$ is reduced (see lower panel of \autoref{fig:projection}), but the dijet search at a hadron collider only depends on $c_3$ and the sensitivity does not change. 
Such linear scaling in the Wilson coefficients is generally robust and the resulting phenomenology does not change significantly, provided the heavy axion is dominated by gluonic decays. 
Overall, our results show the strong and competitive sensitivity on heavy axions at the high-energy muon colliders. 

\section{Heavy QCD Axion Models}
\label{sec:axion_model}
In minimal models, the QCD axion arises as a (pseudo) Nambu-Goldstone boson of a spontaneously broken $U(1)_{\rm PQ}$ symmetry 
at the scale $f_a$. 
This symmetry is anomalous under QCD, which leads to a non-derivative coupling of the QCD axion $a$ to gluons:
\begin{align}
     \mathcal{L}\supset \frac{1}{2}(\partial_\mu a)^2+ \frac{\alpha_s}{8\pi}\left(\bar{\theta} -{a \over f_a} \right)G\widetilde{G}.
\end{align}
Below the QCD confinement scale, this coupling results in a potential for the QCD axion which can be computed using chiral Lagrangian techniques.
At the minimum of this potential $\bar{\theta}=\langle a\rangle/f_a$, which dynamically drives the electric dipole moment of the neutron to a vanishingly small value and solves the strong CP problem.
The axion mass around this minimum is given by~\cite{Weinberg:1977ma, DiVecchia:1980yfw, diCortona:2015ldu}
\es{eq:qcdAxionMass}{
    m^{\rm QCD}_a=\frac{\sqrt{m_um_d}}{m_u+m_d}\frac{m_\pi f_\pi}{f_a}\simeq 5.7 \, {\rm meV}\,\left(\frac{10^{9}{\rm GeV}}{f_a}\right)~.
}
Given this relation and existing astrophysical constraints $f_a\gtrsim 10^8$GeV~\cite{Raffelt:1987yt, ParticleDataGroup:2024cfk}, the maximum axion mass from the QCD contribution is given by $m^{\rm QCD}_a\lesssim  \,$ 60~meV. 

However, the minimal models for the QCD axion are sensitive to Planck-scale symmetry-breaking effects.
Global symmetries, including $U(1)_{\rm PQ}$, are expected to be broken by quantum gravity.
These explicit symmetry-breaking effects are parametrized by including higher-dimensional operators suppressed by the Planck scale in the axion effective field theory: 
\es{}{
V(a) \supset c_n {\Phi^n \over \mpl^{n-4}}+{\rm h.c.} \sim |c_n|\mpl^4 \left({f_a \over \mpl}\right)^n \cos\left({n a \over f_a} +\delta\right),
}
where $c_n\equiv |c_n| e^{i\delta}$ is a complex coefficient with phase $\delta$ and $\Phi \sim f_a \exp(i a/f_a)$ is the complex $U(1)_{\rm PQ}$ scalar with angular mode $a$.
Given the astrophysical constraints $f_a\gtrsim 10^8$GeV~\cite{Raffelt:1987yt, ParticleDataGroup:2024cfk} and $|c_n|\sim 1$, based on bottom-up EFT considerations, all higher-dimensional operators up to $n=8$, would destabilize the axion away from the desired minimum $\langle a\rangle = f_a\bar{\theta}$ where the strong CP problem is solved.
Therefore, these Planck-suppressed contributions spoil the otherwise elegant axion solution, unless all higher-dimensional operators up to $n=8$ are somehow forbidden or suppressed.
Why should the global $U(1)_{\rm PQ}$ symmetry be of such a high quality so that its first violation arises at dimension-9 is the so-called `axion quality problem'~\cite{Ghigna:1992iv, Holman:1992us, Kamionkowski:1992mf, Barr:1992qq}.

Evidently, as $f_a$ becomes smaller, the size of these dangerous Planck-suppressed contributions decrease rapidly.
While $f_a \gtrsim 10^8$~GeV for the traditional QCD axion, {\it heavy} QCD axions can have a much smaller value of $f_a$.
This is because the heavy QCD axions are too heavy to be produced in supernova and stellar environments, and the associated bounds no longer apply for these heavier axions.
This enhancement of the QCD axion mass is due to new contributions to~\eqref{eq:qcdAxionMass} that we will discuss in detail below. Consequently, with $f_a \ll 10^{8}$~GeV, the axion quality problem becomes much less severe, and in some of the parameter space relevant for muon collider probes, it is even absent.
We quantify this in detail below. 
As we have seen in~\autoref{sec:muc_probe}, such heavy QCD axions represent an exciting target at high-energy muon colliders.
To understand what kind of UV theoretical frameworks can give rise to such heavy QCD axions, in this section we consider modifications of minimal axion models~\cite{Kim:1979if,Shifman:1979if,Dine:1981rt,Zhitnitsky:1980tq} that induce significant corrections to the relation~\eqref{eq:qcdAxionMass}. 
In some cases, we will rely on UV modifications of QCD such that the axion receives additional contributions from small instantons. 
These contributions can be estimated either using the dilute instanton gas approximation in the cases where the UV gauge group is weakly coupled, or using the chiral Lagrangian in cases where it is strongly coupled in the UV.
Additionally, for the target $\{f_a,m_a\}$ parameter space, the mechanism that enhances the axion mass does not lead to a new fine-tuning problem for the Higgs mass.\footnote{Note that the UV physics which modifies the axion mass can also lead to corrections to the Higgs potential. 
For muon collider axions with $m_a\sim\,$TeV and $f_a\sim 0.1\,$TeV, these corrections are negligible.
} 
We also discuss how heavy QCD axions arise from ${\mathbb Z}_2$ mirror symmetries. While some aspects of the models we consider below have been previously discussed in the literature, our treatment both serves as a brief review and identifies the specific benchmarks for a muon collider.

\subsection{Product Group $SU(3)^N$}\label{sec:prod_grp}

One way to generate heavy axions is to embed QCD in a semi-simple gauge group in the UV~\cite{Agrawal:2017ksf, Gaillard:2018xgk,Csaki:2019vte}. 
In particular, we consider the scenario where QCD emerges via spontaneous breaking of a product of multiple $SU(3)$'s into their diagonal subgroup, i.e., $ SU(3)_1\times SU(3)_2\times\dots \times SU(3)_N \to SU(3)_c$ at some high scale $M \gg {\rm TeV}$~\cite{Agrawal:2017ksf}.
This is achieved by having multiple Higgs link fields $H_{12}, H_{23}, \cdots, H_{N-1,N}$, where $H_{n, n+1}$ is a bifundamental $(3,\bar{3})$  under $SU(3)_n\times SU(3)_{n+1}$ and a singlet under the rest of the $SU(3)$ gauge groups.
In the UV, the SM quarks are charged only under one $SU(3)$ subgroup, which we choose to be $SU(3)_1$. 
All the $SU(3)_i$ gauge groups have a distinct vacuum angle $\theta_i$ and a corresponding axion $a_i$ is introduced to solve the associated strong CP problem. 
At the scale $M$ where QCD emerges, the QCD coupling $\alpha_s$ is matched the sum of couplings $\alpha_i \equiv g_i^2/(4\pi)$ associated with each of the products groups\footnote{Unlike in \autoref{sec:muc_probe}, in this subsection, we use $\alpha_{1,2}$ to denote the gauge coupling of the $SU(3)_{1,2}$ sectors, not the $U(1)_Y$ and $SU(2)_L$ couplings.}:
\es{eq:matching}{
\frac{1}{\alpha_s(M)}=\sum_{i=1}^N \frac{1}{\alpha_i(M)}.
}
This relation plays a key role in raising the axion mass. 
The small instantons with size $\rho \ll 1/\Lambda_{\rm QCD}$ give a negligible contribution $\propto \exp(-2\pi/\alpha_s(1/\rho))$ within the SM, owing to QCD being asymptotically free ($\alpha_s(1/\rho)\ll 1$).
However, in the product group scenario, the small instantons make a significant contribution.
This is because each $\alpha_i(M) > \alpha_s(M)$, and the small instanton contributions from each $SU(3)_i$ are exponentially enhanced compared to the SM. 
This raises the axion mass.

We assume that each $SU(3)_i$ remains perturbative above the scale $M$, hence allowing a reliable computation of the small instanton contribution to the axion mass.
Below $M$, but above the QCD confinement scale, the EFT containing the $N$ axions is given by
\es{eq:effaction_prodgrp}{
    {\cal L} = \sum_{i=1}^N m_{a_i}^2f_{a_i}^2 \cos\left({a_i \over f_{a_i}} - \bar{\theta}_i\right) + {\alpha_s \over 8\pi} \sum_{i=1}^N \left({a_i \over f_{a_i}} - \bar{\theta}_i\right) G^a_{\mu\nu} \widetilde{G}^{a \mu\nu},
}
where $G^a_{\mu\nu}$ denotes the QCD field strength and $\bar{\theta} = \theta + {\rm arg~det}(M_f)$ includes the appropriate contribution from the phases of the light fermion mass matrix $M_f$.
Since in this model only $SU(3)_1$ has light fermions (below the scale ${\rm min}\{f_{a_i}\}$), $\bar{\theta}_{2,\cdots, N} = \theta_{2,\cdots, N}$.
The QCD strong phase becomes a sum of the individual phases, $\bar{\theta}_{\rm QCD}=\sum_{i=1}^N \bar{\theta}_i$.
The contributions in the first term of Eq.~\eqref{eq:effaction_prodgrp}, proportional to $m_{a_i}^2 f_{a_i}^2$, are due to small instantons from the associated $SU(3)_i$.

These small instanton contributions can be computed using the standard instanton techniques~\cite{tHooft:1976rip} (including the dilute instanton gas approximation~\cite{Callan:1977gz,Callan:1978bm}) which are reliable in this regime since each $SU(3)_i$ is weakly coupled, and we follow the notation described in~\cite{Flynn:1987rs,Csaki:2019vte}.
Given that we are interested in the sensitivity of heavy QCD axions at muon colliders, the relevant parameter space requires some $f_{a_i} \sim {\rm TeV}$ with $M\gg {\rm TeV}$ and therefore $f_{a_i} \ll M$.
This implies the PQ fermions for the $SU(3)_i$ are relevant in generating the associated 't Hooft operator.

\paragraph{One Dirac fermion}

\begin{figure}
    \centering
     \includegraphics[width=0.8\textwidth]{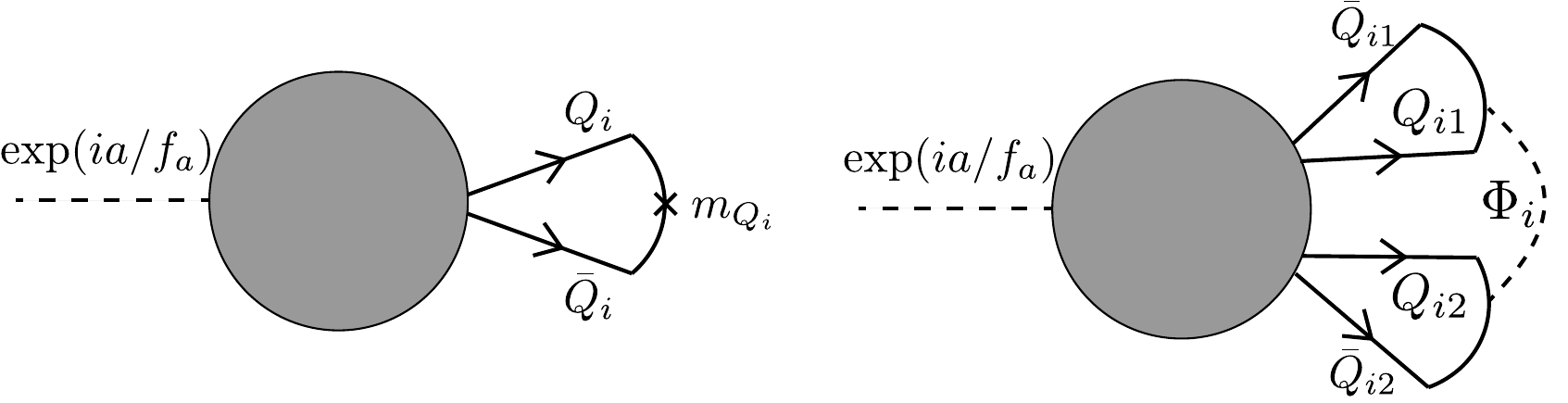}
     \caption{'t Hooft operator contributing to axion mass in product group models with one Dirac fermion (Left) and two Dirac fermions (Right).} 
     \label{fig:inst}
\end{figure}

We first assume there is one family of vector-like fermions $Q_i, \bar{Q}_i$ charged under each $SU(3)_i$, giving rise to the 't Hooft operator shown in \autoref{fig:inst} (left panel). These fermions have a Yukawa coupling $\lambda_i\sim 1$ to the PQ field and obtain a mass $m_{Q_i}=\lambda_if_{a,i}/\sqrt{2}\sim f_{a,i}$. 
For each $SU(3)_i$, the vacuum-to-vacuum amplitude can be estimated as 
\es{eq:W}{
    W_{SU(3)_i} \sim C_{3i} \left(2\pi \over \alpha_i\right)^6 \int \D^4 x \exp\left(i\left( {a_i \over f_{a_i}}-\bar{\theta}_i\right)\right) \left[(N_{f,i}-1)!!\prod_{j=1}^{N_{f,i}}\frac{y_j}{\sqrt{24}\pi}\right] \\\times \int   {\D \rho \over \rho^5} (\Lambda_{SU(3)_i}\rho)^{b_i}  \rho m_{Q_i} \, e^{-S_{i,\rm scalar}}\,,
}
where $b_i = 11 - \frac{1}{6} n_{S_i} - \frac{2}{3}n_{F_i}$ is the one-loop $\beta$-function coefficient for $SU(3)_i$,  with $\beta(g_i)=-b_i g_i^3/(16\pi^2)$ and $n_{S_i}$ ($n_{F_i}$) are the number of scalars (Dirac fermions) in the fundamental representation of $SU(3)_i$. 
The gauge group $SU(3)_i$ becomes strongly coupled at the scale $\Lambda_{SU(3)_i}$.
Finally, the integral $\int \D^4 x$ accounts for the integration over the instanton location.
Note the factor within the square bracket in Eq.~\eqref{eq:W} originates from closing the SM quark legs of the 't Hooft operator via the Higgs loops and hence involves the SM Yukawa couplings $y_i$. 
Thus, $N_{f,1}=6$ and $N_{f,i}=0$ for $i\neq 1$. 
The coefficient $C_{3i}$ is given by~\cite{tHooft:1976snw} 
\es{eq:inst_combinatoric_coeff}{
    C_{3i} = {1\over 2}K_1 e^{-3K_2} e^{-(n_{S_i}-2n_{F_i})\alpha_{1/2}},
}
where $K_1 \approx 0.466$, $K_2\approx 1.678$, and $\alpha_{1/2} \approx 0.146$.
The last factor in Eq.~\eqref{eq:W} results after integrating over the orientation of the instantons within the $SU(3)_i$.
For $SU(3)_1$ and $SU(3)_N$, each of which has only one link field, the scalar action is given by 
\es{}{
    S_{1,\rm scalar} = S_{N,\rm scalar} = 4\pi^2\rho^2 v^2,
}
where $v$ is the VEV of the bifundamental Higgs scalars, while $SU(3)_{i=2,\cdots,N-1}$ has two link fields each, and the scalar action is twice as large. 
The field content charged under $SU(3)_1$, includes one bifundamental scalar link field,  
the SM fermions, and one vector-like PQ fermion, implying $n_{S_1}=3, n_{F_1}=7$, and $b_1 = 35/6$.
Under each of $SU(3)_{2,\cdots,N-1}$, we have two link fields, and one vector-like PQ fermion, implying $n_{S_i}=6, n_{F_i}=1$, and $b_{2,\cdots,N-1} = 28/3$.
Finally, under $SU(3)_N$, we have one scalar link field, and one vector-like PQ fermion, implying $n_{S_N}=3, n_{F_N}=1$, and $b_N = 59/6$.
Note, given the scalar action $S_{i,{\rm scalar}}$, the $\rho$-integral in Eq.~\eqref{eq:W} is convergent in the IR (large $\rho$).

\paragraph{Two Dirac fermions}
Alternatively, we may assume two pairs of vector-like Weyl fermions at each site. 
In this case, the 't-Hooft vertex can be closed via a loop involving the radial component of the PQ field (see the right panel of \autoref{fig:inst}). 
The factor $m_{Q_i}\rho$ in Eq~\eqref{eq:W} is then replaced by the appropriate product of Yukawa couplings (between the vector-like quarks and the radial mode) and a loop factor, which can be computed as $\lambda_1\lambda_2/24\pi^2$ (see, e.g.,~\cite{Csaki:2023ziz}).
For concreteness, we take $\lambda_{1,2}=0.5$ so that the fermions are lighter than the scale $M$.
The $\beta$-function coefficient $b_i$ and $C_{3i}$ also need to be changed accordingly.
The vacuum-to-vacuum persistence amplitude is now given by
\es{eq:W_two}{
    W_{SU(3)_i} \sim C_{3i} \left(2\pi \over \alpha_i\right)^6\int \D^4 x\, e^{i( \frac{a_i}{f_{a_i}}-\bar{\theta}_i)} \left[(N_{f,i}-1)!!\prod_{j=1}^{N_{f,i}}\frac{y_j}{\sqrt{24} \pi}\right]\frac{\lambda_1 \lambda_2}{24 \pi^2}\int  {\D \rho \over \rho^5} (\Lambda_{SU(3)_i}\rho)^{b_i}   e^{-S_{i,\rm scalar}} 
    ,\\
    \sim  \left({C_{3i}\lambda_1 \lambda_2 \over 24\pi^2}\right)\left(2\pi \over \alpha_i\right)^6 \int \D^4 x\, e^{i( \frac{a_i}{f_{a_i}}-\bar{\theta}_i)}\left[(N_{f,i}-1)!!\prod_{j=1}^{N_{f,i}}\frac{y_j}{\sqrt{24} \pi}\right] \Lambda_{SU(3)_i}^{b_i} \left({1\over 2\pi v_i}\right)^{b_i-4} 
    \Gamma\left({b_i \over 2}-2 \right)\,,
 } 
where $v_i=v$ for $i=1, N$, and $v_i = \sqrt{2}v$ for $i=2,\cdots,N-1$.
Due to the appearance of $m_{Q_i}$ in \eqref{eq:W}, the axion mass for the one Dirac fermion case is smaller than that for two Dirac fermions.
Therefore, for concreteness, we will consider only the two Dirac fermion case below for numerical benchmarks.
The axion potential is then derived using $\int \D^4 x \,V(a_i) = - (W_{SU(3)_i}+\bar{W}_{SU(3)_i})$,
\es{eq:}{
V(a_i) = -C_{3i} (N_{f,i}-1)!!\left(\prod_{j=1}^{N_{f,i}}\frac{y_j}{\sqrt{24} \pi}\right) \left(2\pi \over \alpha_i(M)\right)^6 \frac{\lambda_1\lambda_2 M^4}{24 \pi^2} \left({M\over 2\pi v_i}\right)^{b_i-4}\\ \times \Gamma(b_i/2-2 ) e^{-\frac{2\pi}{\alpha_i(M)}}\cos(\frac{a_i}{f_{a_i}}-\bar{\theta}_i).
}
The chiral suppression factor for $i=1$ is given by
\es{eq:ch_sup}{
K = (N_{f,1}-1)!!\prod_{j=1}^{N_{f,1}}\frac{y_j}{\sqrt{24} \pi} = {40\over 9}{y_u y_d y_c y_s y_b y_t \over (16 \pi^2)^3}.
}
Thus, collecting all the factors for the $N=3$ case\footnote{Note that for $N=2$, there is only one heavy QCD axion within the reach of a muon collider. However, the other axion is much lighter and ruled out by astrophysical constraints.} and assuming $v\simeq M$, we obtain
 \es{eq:ax_mass_final}{
    m_{a,i}^2f_{a,i}^2\simeq    \gamma_{i}  \left(2\pi \over \alpha_i\right)^6   M^4  \exp\left(-\frac{2\pi}{\alpha_i(M)}\right)\,,
}
where $\gamma_1=1.4\times 10^{-28},\gamma_2=1.1\times 10^{-10},\gamma_3=3.9\times 10^{-10}$ and we have used $b_1=31/6, b_2=26/3$, and $ b_3=55/6$. 

\begin{figure}
     \includegraphics[width=0.6\linewidth]{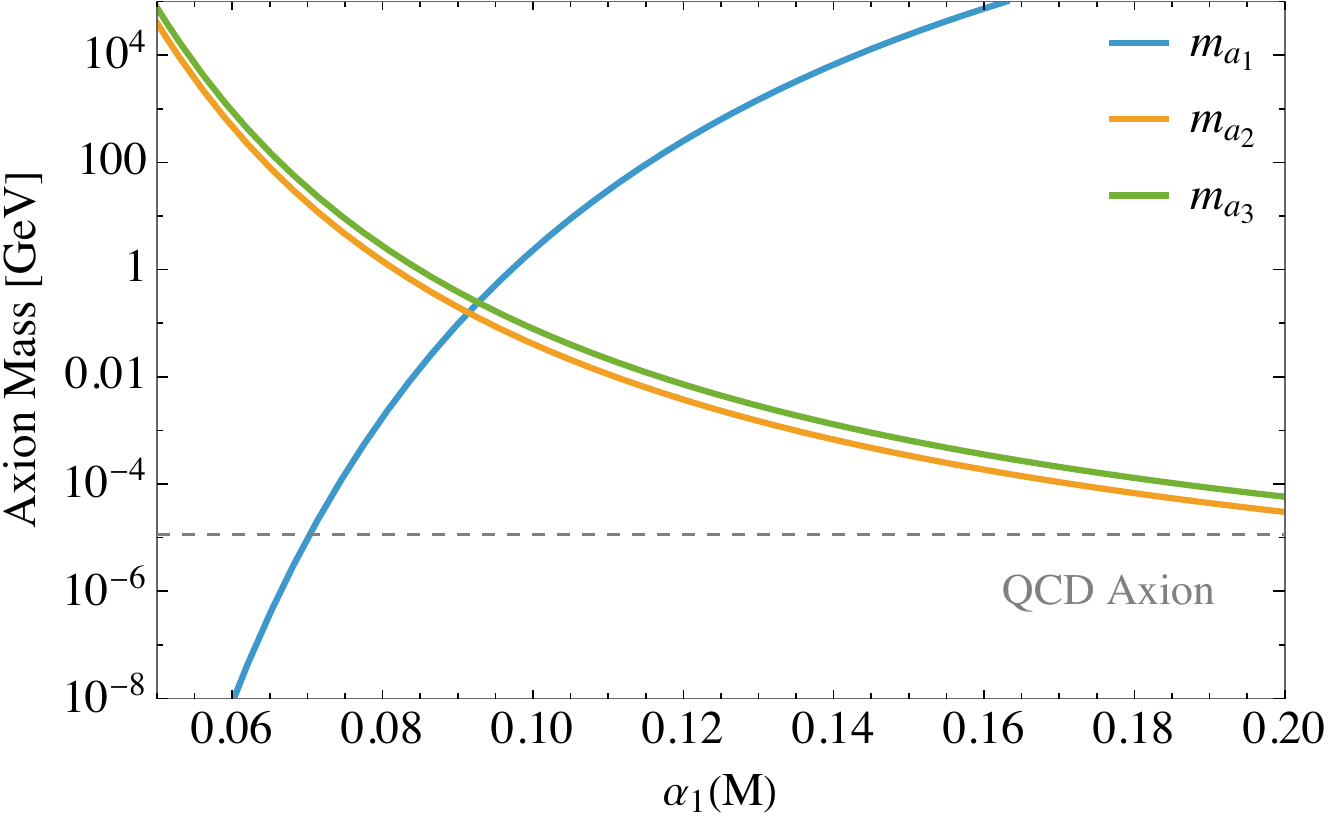} 
    \caption{The axion mass in the $SU(3)^3$ model as a function of $\alpha_1(M)$. As a benchmark for $f_{a,2}=f_{a,3}=M=5\times 10^{12}$~GeV, $f_{a,1} = 500$~GeV, $\alpha_1(M)=0.12$, the axion masses are $m_{a_1} \approx 250$~GeV, $m_{a_2}\approx 4$~MeV, and $m_{a_3} \approx 7$~MeV. We have chosen $\alpha_2(M) = \alpha_3(M)$ such that~\eqref{eq:matching} is obeyed. The horizontal dashed line denotes the conventional QCD axion mass value for this choice of $f_a$ using~\eqref{eq:qcdAxionMass}. 
    }
    \label{fig:ax_mass_SU(3)3}
\end{figure}

The field that couples to $G\widetilde{G}$ is given by a linear combination of all the axions as in~\eqref{eq:effaction_prodgrp}.
For $N=3$, there are three such axions whose masses depend sensitively on $\alpha_i$ via~\eqref{eq:ax_mass_final} with $\alpha_i$ obeying~\eqref{eq:matching}.
In \autoref{fig:ax_mass_SU(3)3} we show how the masses of the three axions vary as a function of $\alpha_1(M)$ (we fix $\alpha_2(M) = \alpha_3(M)$ and $f_{a,2}=f_{a,3}=M$ for simplicity).\footnote{We note that the second term in \eqref{eq:effaction_prodgrp} generates a mass mixing between the different axions. The QCD contribution to the topological susceptibility, $\chi_{\rm QCD}$ generates an off-diagonal term in the mass squared matrix given by $ {\chi_{\rm QCD}/f_{a_1}f_{a_2}}$. But since we work in the regime where the conventional QCD axion mass $\sim\sqrt{\chi_{\rm QCD}}/f_{a_1}\ll m_{a_1}$ and $f_{a_1}\ll f_{a_2}$, the mass mixing term ${(\chi_{\rm QCD}/f_{a_1}^2)(f_{a_1}/f_{a_2}})\ll m_{a_1}$ and hence is negligible. }
We see that for certain values of $ \alpha_1(M)$, some of the axions can be significantly lighter and hence constrained by existing astrophysical and cosmological constraints.
While a detailed study of these constraints is beyond the scope of the present work, we focus on scenarios such that $a_1$ is heavy enough to avoid astrophysical and cosmological constraints, while light enough to be probed at a muon collider.
On the other hand, $a_{2,3}$ are light but too weakly coupled to be probed.
So far, the scale $M$ has been treated as a free parameter.
However, there are restrictions due to possible misalignment due to higher-dimensional CP-odd operators $\mathcal{O}(1/M_{\rm Pl}^2)$, as well as non-perturbative effects assumed to be generated at the Planck scale~\cite{Dine:1986bg,Bedi:2022qrd,Csaki:2023ziz,Bedi:2024wqg, Agrawal:2017evu}. 
These operators can shift $\bar\theta_{\rm QCD}$, which is constrained to be $\lesssim10^{-10}$.  This shift can be estimated to be $(2\pi M)^2/M_{\rm Pl}^2$, implying $M\lesssim 10^{13}$GeV. 

Consistent with this restriction, in \autoref{fig:final_plot}, we fix $M=5\times 10^{12}$~GeV and $\alpha_2(M)\approx \alpha_3(M)$, which implies $m_{a_2}\approx m_{a_3}$.
We then show how $m_{a_1}$ varies with $f_a$ (labeled `Product Group') and observe that a significant range of the parameter space can be probed at a muon collider.
Additional parameter space can be obtained by choosing different values of $M$ and $\alpha_1(M)$, as detailed in \autoref{fig:summary} (top left).
 
Since the axion mass enhancement relies on coupling a set of fundamental axions to the product group $SU(3)_1\times SU(3)_2\times SU(3)_3$, the heavy QCD axions do not necessarily couple to photons or the $W$ and $Z$ bosons directly. 
However, $SU(2)_L$ and $U(1)_Y$ couplings can arise if, for example, the heavy PQ fermions are charged under $SU(2)_L$ and $U(1)_Y$.
To make our analysis general, we will also assume $SU(2)_L$ and $U(1)_Y$ couplings of heavy QCD axions.

\subsection{Extra Dimension}\label{sec:extradim}

\begin{figure}
     \includegraphics[width=0.6\linewidth]{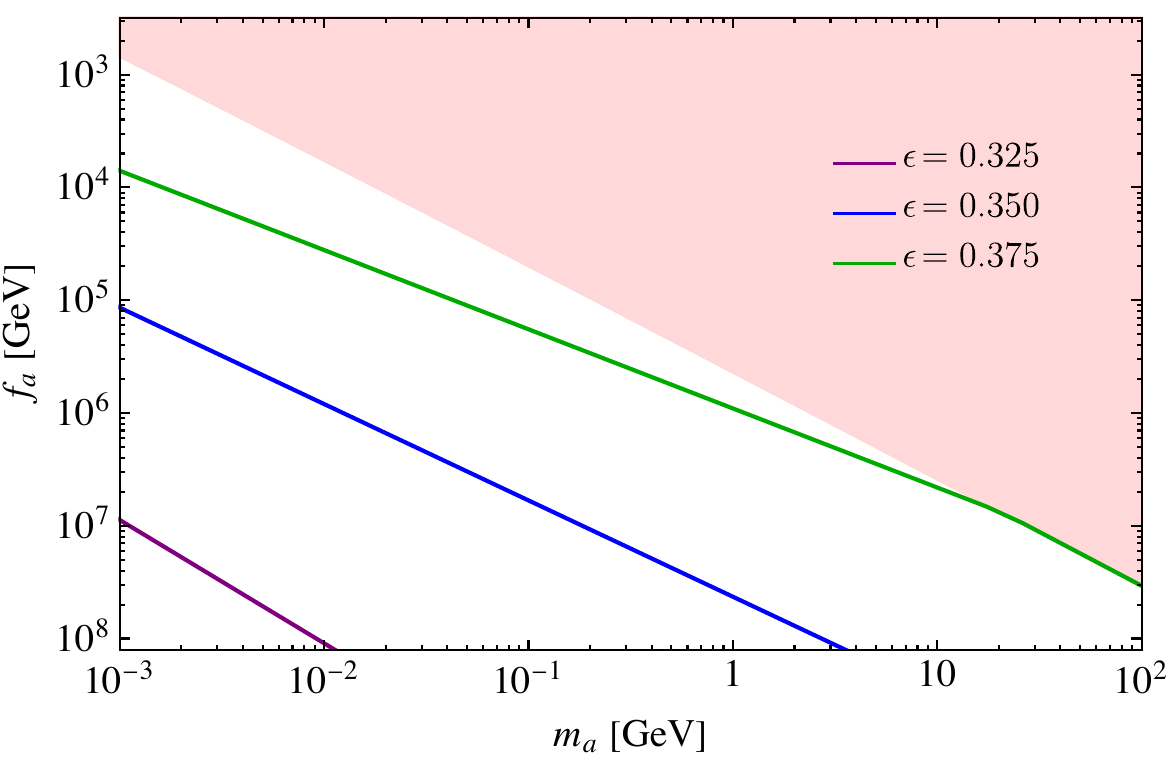} 
     \caption{
      Dependence of the axion decay constant with axion mass, when QCD propagates in a flat extra dimension and the axion is identified with the scalar component of a 5D gauge field (with $b_{\rm CS}=1,\,c_3=1$). The red shaded region corresponds to the region in which 5D perturbativity breaks down and the axion mass is given by~\eqref{eq:max_5d_Axion_mass}. To the left of the kink on the green line, the axion mass is given by~\eqref{eq:5d_axion_mass} where the instanton gas approximation is valid, while to the right, it is determined by~\eqref{eq:max_5d_Axion_mass}.} 
    \label{fig:ax_mass_5dGauge}
\end{figure}

\paragraph{Axion from bulk gauge field}
We now consider scenarios where instead of having multiple $SU(3)$ gauge groups in the UV, we have a single $SU(3)$ gauge group throughout, giving the standard QCD running at low energies.
However, the running is modified in the UV such that $\alpha_s$ becomes larger in the UV, and the small instanton contributions raise the axion mass~\cite{Holdom:1982ex, Holdom:1985vx, Flynn:1987rs}.
This effect can be computed in a 5D UV modification where QCD propagates in the bulk of a flat extra dimension of size $L=\pi R$ with orbifold boundary conditions~\cite{Gherghetta:2020keg}. 
In this construction, QCD is part of a bulk $SU(3)$ gauge field $A_{M}$ with a coupling $g_5$ and a field strength $G_{MN}$, while the axion is part of a bulk $U(1)$ gauge field $B_{M}$ with a field strength $F_{MN}$.
For simplicity, we also choose the gauge coupling of the $U(1)$ group to be $g_5$.
By choosing appropriate boundary conditions, it can be ensured that the only zero modes of $A_M$ and $B_M$ are the QCD gauge field $A_\mu^{(0)}$ and the axion $B_5^{(0)}$, respectively.
The axion coupling to gluons is then generated via a 5D Chern-Simons term, with a dimensionless coefficient $b_{\rm CS}$: 
\begin{align}
    {S}\supset -\int_0^L \D y\int \D^4x\left\{\frac{1}{2 g_5^2} {\rm tr\,}[G_{MN}G^{MN}]+\frac{1}{4g_5^2}
    F_{MN}F^{MN} -\frac{b_{\rm CS}}{32\pi^2}\epsilon^{MNRST}B_M{\rm tr}[G_{NR}G_{ST}]\right\}~,\label{eq:5dAction}
\end{align}
where $y$ is the extra dimensional coordinate and the Latin indices run over the coordinates $x$ and $y$.
Imposing the boundary conditions $A_5=\partial_5 A_{\mu}=0$ and $B_\mu=\partial_5 B_5=0$ at the two boundaries, performing a KK decomposition, and canonically normalizing the kinetic term, we obtain the low-energy 4D Lagrangian,
\begin{align}
    \mathcal{L}\supset -\frac{1}{2}{\rm tr\,}[G_{\mu\nu}G^{\mu\nu}]+\frac{1}{2}(\partial_\mu a)^2+\frac{b_{\rm CS}L g_s^3}{32\pi^2}a\,{\rm tr\,}G_{\mu\nu}\widetilde{G}^{\mu\nu}~,
    \label{eq:5dModel_4Daction}
\end{align}
where we have performed the transformation $A_{\mu}\to g_sA_{\mu}$, and redefined $G_{\mu\nu}$ accordingly. The axion is identified as $a(x)=B_5^{(0)}(x)/g_s$, which also determines $f_a/c_3=1/(\pi g_s b_{\rm CS}R)$, 
and $g_s=g_5/\sqrt{\pi R}$. 
The contributions due to the Kaluza-Klein modes modify the running of the effective gauge coupling above the compactification scale $1/R$. The effective action then becomes~\cite{Gherghetta:2020keg}
\begin{equation}
    S_{\rm eff}\simeq   \dfrac{8\pi^2}{g^2(1/R)}-N \xi(R/\rho)\frac{R}{\rho}+b_0\ln\frac{R}{\rho}~,
    \label{eq:5D_eff_action}
\end{equation}
where $b_0$ is the $\beta$-function coefficient of QCD at the scale $1/R$ and $N=3$ for the case of $SU(3)$.
The function $\xi(R/\rho)$ is given in~\cite{Gherghetta:2020keg} and is approximately $\xi \sim 1/3$ for $R/\rho \gtrsim 20$.
The term given by $\xi$ originates from the positive frequency modes of Kaluza-Klein gluons and since $\xi>0$, this contribution reduces $S_{\rm eff}$ and effectively raises the axion mass.
The 5D theory has a naive dimensional analysis (NDA) cutoff given by 
\begin{align}
    \Lambda_5^{\rm NDA}
   \sim \left(\frac{g_5^2}{24\pi^3}\right)^{-1}= \frac{6\pi}{\alpha_s }\frac{1}{R}~,
\end{align} which corresponds to the scale at which the 5D gauge theory becomes strongly coupled.
The instanton integral is carried out for $1/R < 1/\rho < \Lambda_5$.
To relate the cutoff $\Lambda_5$ to $\Lambda_5^{\rm NDA}$, it is helpful to define a perturbativity parameter $\epsilon$ via
\begin{equation}\label{eq:5d_prt_lmt}
    \Lambda_5 R=\frac{6\pi\epsilon}{\alpha_s(1/R)}~,
\end{equation} 
so that $\epsilon < 1$ ensures $\Lambda_5 < \Lambda_5^{\rm NDA}$, necessary for perturbativity.
We further need to impose this perturbativity condition for all $R^{-1}<\rho^{-1}<\Lambda_5$ when the appropriate running of the gauge coupling \eqref{eq:5D_eff_action} is taken into account. 
When the theory is strongly-coupled, i.e., the parameter $\epsilon$ is too large, we can no longer use the dilute instanton gas approximation. 
The maximal mass of the axion (for a given $f_a$) in this scenario can be estimated as, 
\begin{align}
    m_a^2{f_a^2}\simeq \left(\prod_{j=1}^6\frac{y_j}{4 \pi}\right)  \, {\Lambda_5^4} = 10^{-23}\, {\Lambda_5^4}~.
    \label{eq:max_5d_Axion_mass}
\end{align}
This determines an upper limit on $\epsilon$.
In the dilute instanton gas approximation, using the relation between the decay constant and the scale of the extra dimension, we can obtain the axion potential,
\es{eq:ax_pot_ed}{
V(a) =& -C_{3} K \left(\frac{2\pi}{\alpha_s(1/R)}\right)^6 2 \cos(a/f_a-\bar{\theta}) \int {\D \rho \over \rho^5} e^{-S_{\rm eff}},\\
=& -C_{3} K \left(\frac{2\pi}{\alpha_s(1/R)}\right)^6 2 \cos\left({a \over f_a}-\bar{\theta}\right) \exp\left({-{2\pi \over \alpha_s(1/R)}}\right) (\Lambda_5 R)^{3-b_0}\exp(\Lambda_5 R)R^{-4}, 
}
for $\Lambda_5 R \gg 1$.
This determines the axion mass to be
\es{eq:5d_axion_mass}{
m_a  \simeq & f_a\sqrt{2\,K\,C_{3}}\left(\frac{2\pi}{\alpha_s(1/R)}\right)^3 \exp\left({-\frac{\pi}{\alpha_s(1/R)}+{3\pi \epsilon \over \alpha_s(1/R)}}\right)\left({6\pi \epsilon \over \alpha_s(1/R)}\right)^{(3-b_0)/2} (\pi g_s b_{\rm CS}/c_3)^2\,,\\  
}
where $1/\alpha_s(1/R) =- b_0/(2\pi)  \ln(R\Lambda_{\rm QCD})$ and $C_3$, $K$ are defined in \eqref{eq:inst_combinatoric_coeff} and \eqref{eq:ch_sup}, respectively. 
By ignoring the $\mathcal{O}(1)$ log corrections coming from the $(2\pi/\alpha_s)^6$ factor in \eqref{eq:5d_axion_mass}, the axion mass expression can be more simply written as
\es{eq:5d_gauge_axion}{
\left(\frac{m_a}{34\,{\rm MeV}} \right) \simeq &\,(5.5\times 10^{25})^{{\epsilon\over 0.38}-1}\left(\frac{\epsilon}{0.38}\right)^{-2}\left(\frac{f_a}{50\,{\rm TeV}}\right)^{(21\epsilon-5)/2} \left(\frac{b_{\rm CS}}{c_3}\right)^{(21\epsilon-3)/2}\,,
}
where we have used $b_0=7$. This expression reveals that the axion mass is too light to be relevant at the MuC. This is confirmed by plotting the variation of $m_a$ with $f_a$ using  \eqref{eq:5d_axion_mass} in \autoref{fig:ax_mass_5dGauge} where the parameter space relevant for a muon collider lies in the red-shaded, non-perturbative regime.  
Consequently, we now turn to a modification of the 5D model in Eq.\eqref{eq:5dAction}.

\paragraph{Axion from a bulk scalar}
An extra-dimensional axion target for a muon collider can nonetheless arise if we assume the axion originates not from a gauge field as in \eqref{eq:5dAction}, but instead from a 5D PQ scalar field, $\Phi(x,y)$~\cite{Dienes2000}~\footnote{We may also consider $\Phi$ to be a 4D scalar localized on the boundary, although we do not pursue this possibility here.}
with an action given by
\begin{align}
    S_5 \supset \int_0^{\pi R} \D y\int \D^4x\, M_5\,\left( |\partial\Phi|^2
    -\frac{\lambda}{2}\left( |\Phi|^2-\frac{\hat{f_a}^2}{2}\right)^2
     \right)~,
     \label{eq:S5bulkscalar}
\end{align}
where $M_5$ is a cutoff scale of the 5D theory, such as $ \Lambda_5^{\rm NDA}$.
Eventually, $\Phi$ obtains a VEV at the scale $\hat{f_a}$, and hence can be parameterized as $\Phi(x,y)\simeq \frac{\hat f_a}{\sqrt{2}}e^{ia_5(x,y)/\hat f_a}$ where the angular component $a_5(x,y)$ is the 5D axion field. 
The effective action for $a_5$ is then given by
\begin{align}
    S \supset \int  \D y \, \D^4x\,   \left( 
   \frac{1}{2}M_5(\partial_\mu a_5)^2+\frac{\alpha_s}{8\pi}\frac{a_5}{\hat f_a}G_{\mu\nu}\tilde G^{\mu\nu}\delta(y)\right)~,
    \label{eq:5d_axion_GG}
\end{align}
where we have introduced a boundary coupling of the 5D axion field to the 5D gluons. We limit ourselves to the (massless) zero mode of the 5D axion field, which is the Nambu-Goldstone boson from the 4D viewpoint and hence identified as the 4D axion $a(x)$, up to an overall normalization constant. Upon integrating out the extra dimension, we obtain an effective 4D decay constant given by $f_a=\hat f_a \sqrt{M_5\pi R}$.\footnote{This volume effect was noted in Ref.~\cite{Dienes2000}, which implies that the effective $f_a$ can be parametrically larger than the 5D symmetry breaking scale, $\hat f_a$.} 
Using the effective action for QCD in the bulk \eqref{eq:5D_eff_action}, we can again employ the potential given by \eqref{eq:ax_pot_ed}. But unlike the case in which the axion arises from a bulk gauge field, apart from the size of the fifth dimension, $f_a$ is also related to an independent cutoff scale $M_5$, and hence can be treated as an independent parameter.
This gives rise to heavier axions compared to the previous case and these heavier axions can be probed at muon colliders.

As in section \ref{sec:prod_grp}, we supplement the model with a KSVZ vector-like fermion. However, we assume the KSVZ fermion is on the $y=0$ boundary  in order to obtain the maximum axion mass enhancement (see appendix A.1 of Ref.~\cite{Gherghetta:2020keg}), consistent with the boundary coupling in \eqref{eq:5d_axion_GG}. We assume that the PQ-charged fermion has an order one Yukawa coupling to $\Phi$ on the boundary, leading to a mass $\sim f_a$. This modifies \eqref{eq:5d_axion_mass} and the axion potential is given by
\es{eq:ax_pot_ed_2}{
V(a) =& -C_{3} K \left(\frac{2\pi}{\alpha_s(1/R)}\right)^6 2 \cos(a/f_a-\bar{\theta}) \int {\D \rho \over \rho^5} (m_Q \rho) e^{-S_{\rm eff}}.
}
This leads to the relation 
\es{}{
    m_a^2f_a^2  \simeq & \,2\,K \,e^{0.292\cdot 7}C_3\left(\frac{2\pi}{\alpha_s(R^{-1})}\right)^6 e^{-\frac{2\pi}{\alpha_s(R^{-1})}+\Lambda_5R}(\Lambda_5R)^{2-b_0}f_a\,R^{-3},
}
and an axion mass
\begin{align}
 \left(\frac{m_a}{300\,{\rm GeV}} \right) 
      \simeq &\,(4.5\times 10^{31})^{{\epsilon\over 0.38}-1}\left(\frac{\epsilon}{0.38}\right)^{-13/6}\left(\frac{f_a}{{\rm TeV}}\right)^{-1/2} \left(\frac{R^{-1}}{10^7\,{\rm GeV}}\right)^{19\epsilon-29/6}~,\label{eq:5d_scalar_mass1}
\end{align}
where we have again ignored the $\mathcal{O}(1)$ logarithmn  arising from the $(2\pi/\alpha_s)^6$ factor in \eqref{eq:ax_pot_ed_2}.

\begin{figure}
    \centering
    \includegraphics[width=0.49\linewidth]{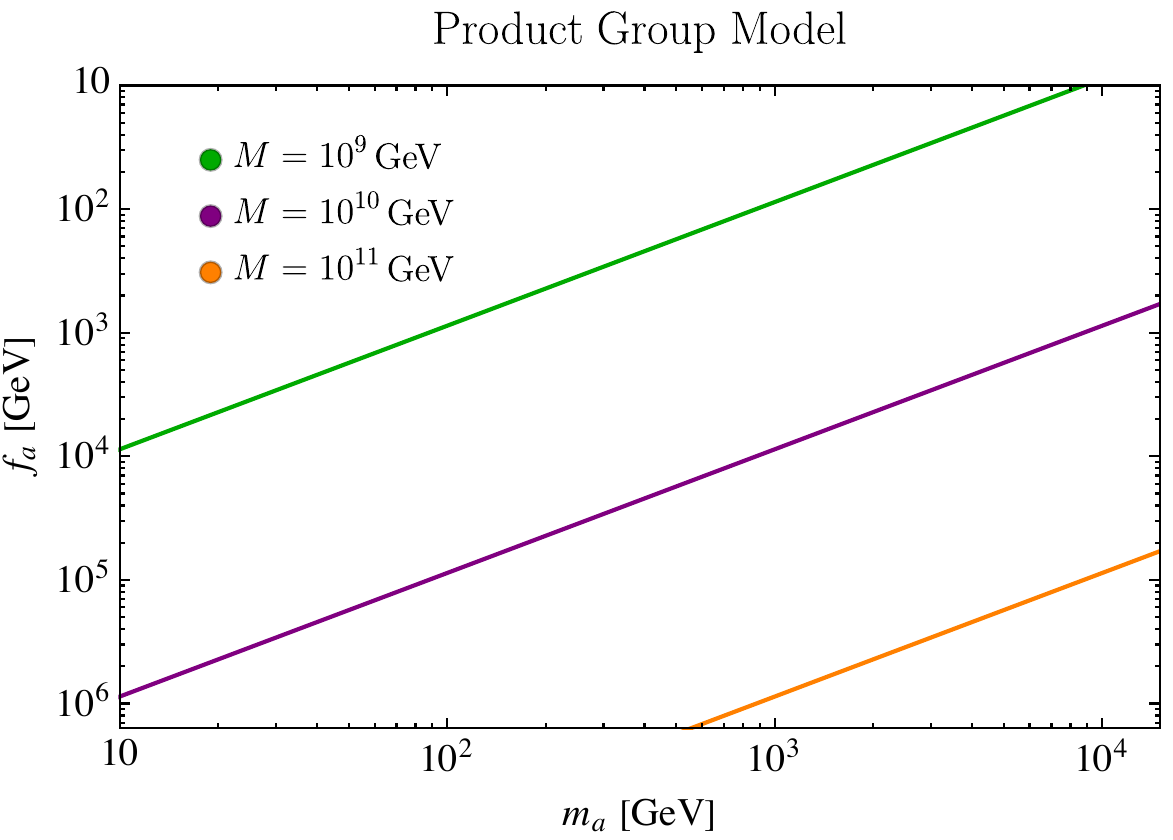}
    \includegraphics[width=0.49\linewidth]{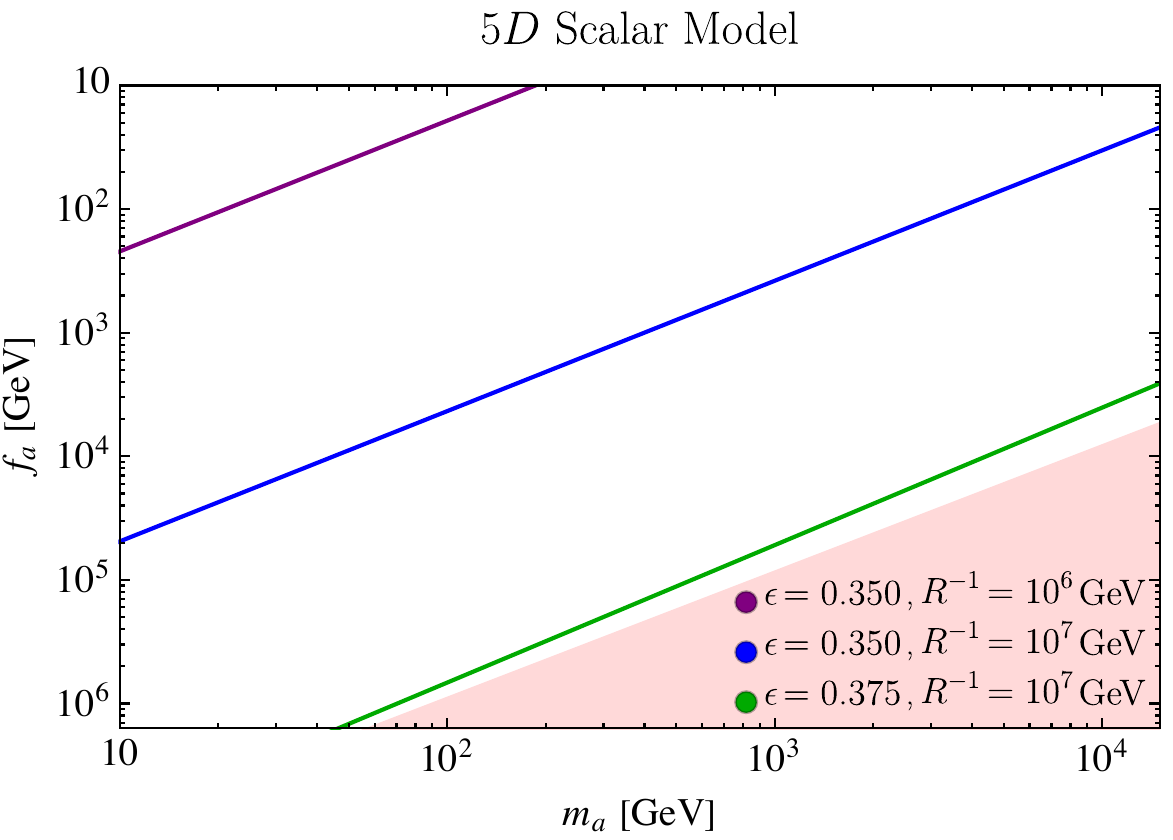}
    \includegraphics[width=0.49\linewidth]{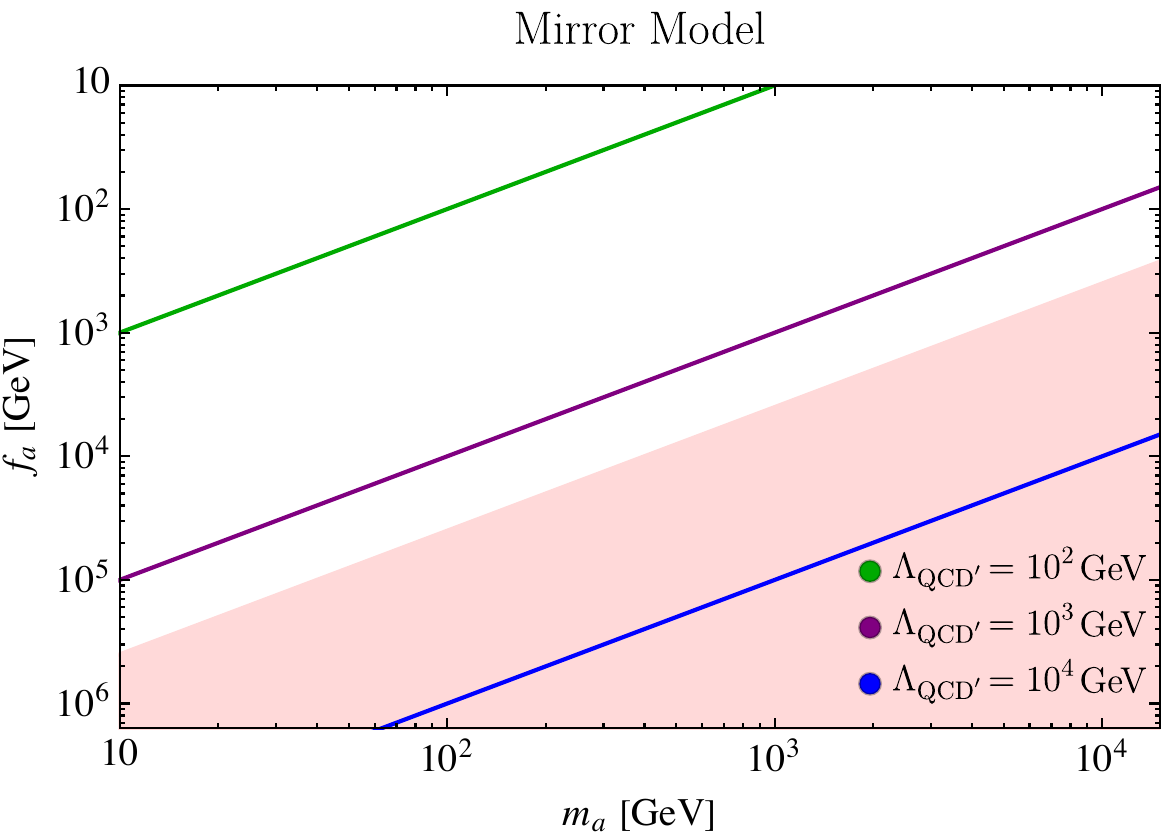} \includegraphics[width=0.49\linewidth]{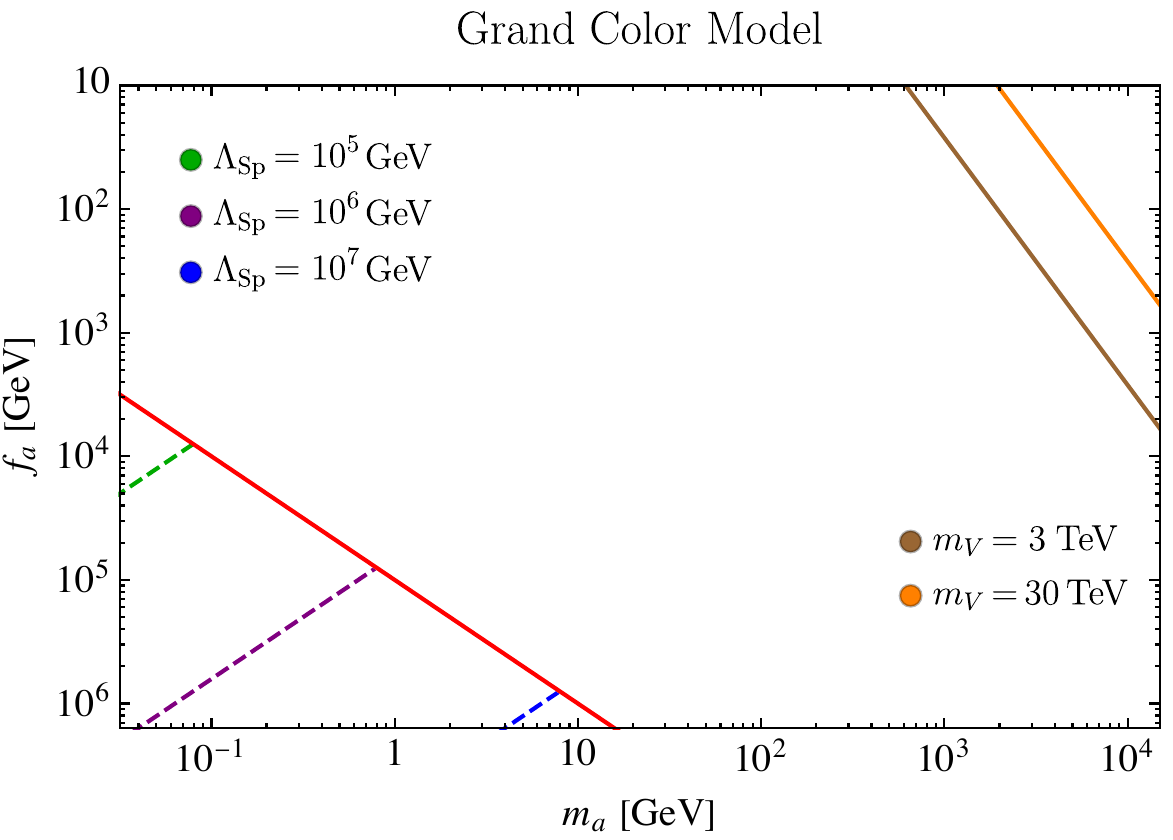}
  \caption{Relation between the axion mass $m_a$ and the decay constant $f_a$ for different heavy QCD axion models.
  {\it Top left}: 
  Axion mass for the product-group model (\autoref{sec:prod_grp}).
    To derive the lines for each value of $M$, we fix $\alpha_i(M)$ such that~\eqref{eq:matching} is obeyed with $\alpha_2(M)=\alpha_3(M)$, and plot the mass of the heaviest axion. {\it Top right}: The axion mass arising from a bulk scalar field (\autoref{sec:extradim}), coupled to QCD propagating in the bulk, for two KSVZ fermions as given by \eqref{eq:5d_scalar_mass2}. 
    If  $f_a$ is too large, the instanton gas approximation breaks down and the axion mass is given by \eqref{eq:max_5d_Axion_mass}--shown by the red shaded region for $R^{-1}=10^7~$GeV.  {\it Bottom left}: The axion mass arising from the mirror ($\mathbb{Z}_2$) models (\autoref{sec:mirror}) for various values of $\Lambda_{\rm QCD^\prime} $ using~\eqref{eq:mirror_mass}. The shaded region corresponds to the bound $\Lambda_{\rm QCD^\prime} \lesssim 2$~TeV due to the constraint on the mirror Higgs VEV. {\it Bottom right}: The axion mass in color unification models (\autoref{sec:colorunif}). The dashed lines show the mass for the grand color axion \eqref{eq:GC0} and the solid lines correspond to the model with a massless up quark and massive vector-like quarks~\eqref{eq:GCmass2}. 
}
    \label{fig:summary}
\end{figure}

Alternatively, we may assume the UV completion contains two KSVZ vector-like fermions with $\mathcal{O}(1)$ Yukawa couplings to the PQ scalar field, $\Phi$. We then replace the factor $m_Q\rho$ in \eqref{eq:ax_pot_ed_2} by $1/12\pi^2$, together with the corresponding changes in $C_3$ and the $\beta$-function coefficient to obtain 
\begin{align}
\left(\frac{m_a}{285\,{\rm GeV}} \right) 
      \simeq &\,(8.1\times 10^{26})^{{\epsilon\over 0.38}-1}\left(\frac{\epsilon}{0.38}\right)^{-4/3}\left(\frac{f_a}{{\rm TeV}}\right)^{-1} \left(\frac{R^{-1}}{10^6\,{\rm GeV}}\right)^{17\epsilon-11/3}~.\label{eq:5d_scalar_mass2}
\end{align}
Compared to \eqref{eq:5d_gauge_axion}, this expression shows that the axion mass has values that can be probed at a muon collider.
The scale $\Lambda_5$ (see \eqref{eq:5d_prt_lmt}) is further restricted by the possible misalignment due to CP-violating higher-dimensional operators~\cite{Dine:1986bg,Bedi:2022qrd,Csaki:2023ziz,Bedi:2024wqg}.  This constrains $\Lambda_5^2/\mpl^2\lesssim10^{-10}$, implying $\Lambda_5\lesssim 10^{14}$~GeV and thus $m_a\lesssim10^{17} {\rm GeV}^2/f_a$, based on~\eqref{eq:max_5d_Axion_mass}. The variation of $m_a$ with $f_a$ is shown in \autoref{fig:summary} (top right), where the values $m_a\sim$~TeV and $f_a\sim$~TeV are readily attainable for an extra dimension size $R^{-1} \sim 10^{5-7}$~GeV. Note that an effective decay constant $f_a\sim$~TeV can be obtained by choosing a 5D cutoff scale $M_5\sim 10^2 R^{-1}$ and a 5D symmetry breaking scale $\hat f_a\sim 60\,$GeV in the Lagrangian \eqref{eq:S5bulkscalar}.

Since the electroweak gauge bosons are assumed to be localized on the boundary, 
the axion does not necessarily couple to them at tree level 
and $c_{1,2}$ receive contributions only due to loop corrections, similar to the product group models.  
If such couplings are indeed introduced as brane terms with order one couplings, the 4D effective couplings are suppressed by the scale associated with the size of the extra dimension, given by $\sqrt{M_5\pi R}$, similar to the axion coupling to gluons, implying $c_{1,2}\sim c_3$.

\subsection{Mirror Symmetry}
\label{sec:mirror}
Another class of models that enhances the axion mass introduces a ``mirror" copy of the SM, with exactly the same field content~\cite{Rubakov:1997vp,Berezhiani:2000gh,Hook:2014cda,Fukuda:2015ana,Hook:2019qoh}. 
All the parameters in both sectors, including the $\bar{\theta}$ parameters, are assumed to be $\mathbb{Z}_2$ symmetric in the far UV, under the exchange of the SM with its mirror copy, SM$^\prime$.
This $\mathbb{Z}_2$ symmetry is softly broken by the VEV, $v_{\rm SM^\prime}$ of the mirror Higgs where we assume $v_{\rm SM^\prime} \gg v_{\rm SM}$, the VEV for the SM Higgs.
The hierarchy $v_{\rm SM^\prime} \gg v_{\rm SM}$ implies that the mirror quarks are accordingly heavier than the SM quarks, and this results in a parametrically larger confinement scale for the mirror $SU(3)$.
We will focus on the following hierarchy
\begin{align}
 \Lambda_{\rm QCD} \ll  v_{\rm SM} \lesssim \Lambda_{\rm QCD^\prime}  \ll v_{\rm SM^\prime}~.
\end{align}
When the model is supplemented with a KSVZ-type axion to solve the strong CP problem, the axion receives a mass contribution from mirror QCD, in addition to the standard QCD contribution, given by~\cite{Berezhiani:2000gh,Hook:2014cda,Fukuda:2015ana,Hook:2019qoh}
\begin{align}
    m_a \simeq (75\, {\rm MeV})^2\frac{1}{ f_a} \left( \frac{v_{\rm SM^\prime}}{v_{\rm SM}} \right)^{1/2}\left(
    \frac{\Lambda_{\rm QCD^\prime} }{ \Lambda_{\rm QCD}}\right)^{3/2}~,
\end{align}
for $\Lambda_{\rm QCD^\prime} \gtrsim m_{u^\prime} \approx 10^{-5}v_{\rm SM^\prime}$, which corresponds to the hierarchy between the mirror up quark mass $m_{u^\prime}$ and the mirror QCD confinement scale $\Lambda_{\rm QCD^\prime}$. In the case when $\Lambda_{\rm QCD^\prime} \lesssim m_{u^\prime}$, we instead have
\begin{align}
    m_a \simeq  \frac{1}{ f_a} \Lambda_{\rm QCD^\prime} ^2\,.
    \label{eq:mirror_mass}
\end{align}
We show the $\{f_a,m_a\}$ parameter space in \autoref{fig:summary} (bottom left). As in previous models, CP-odd higher-dimensional operators~\cite{Dine:1986bg,Bedi:2022qrd,Csaki:2023ziz,Bedi:2024wqg} imply $\Lambda_{\rm QCD^\prime} \lesssim 10^{14}$GeV, and hence 
$m_a\lesssim 10^{14}{\rm GeV} (10^{14}{\rm GeV}/f_a)$.
However, a much stronger bound originates from dimension-six operators such as 
\begin{equation}
    \frac{\alpha_s}{8\pi} \frac{1}{\mpl^2}|H'|^2 G'\widetilde{G'}\,,
\end{equation} 
which requires $v_{\rm SM^\prime} \lesssim 10^{14}$~GeV. 
Otherwise, $\bar{\theta}'\gtrsim 10^{-10}$ and the strong CP problem would reappear. This puts an upper bound on the mirror confinement scale, $\Lambda_{\rm QCD^\prime} \lesssim 2$~TeV.\footnote{Alternatively, the mirror $SU(3)$ gauge group can be spontaneously broken~\cite{Co:2024bme} so that it does not confine.
In this case, the relation~\eqref{eq:mirror_mass} is modified and depends on the mirror $SU(3)$ symmetry breaking scale. This allows for a different flexibility in the $\{f_a,m_a\}$ parameter space.
}
The parameter space for this model is depicted in \autoref{fig:summary} (bottom left).

In contrast to the previous models, which modify the QCD gauge group in the UV, the $\mathbb{Z}_2$ symmetry requires a copy of all the SM fields, including the photons and the $W,Z$ bosons.
Hence, it is natural to have $c_{1,2}\sim 1$ and this will increase the production rates, as shown in \autoref{fig:projection}.
 
\subsection{Color Unification}
\label{sec:colorunif}

A distinct class of models that have heavy axions rely on a UV modification of the QCD gauge group into a larger group~\cite{Gherghetta:2016fhp, Valenti:2022tsc,Bedi:2024kxe}, such as $ SU(2N+3)$ that is Higgsed down to $SU(3)_c$: 
\begin{equation}
SU(2N+3)\times U(1)_{Y'}\rightarrow  Sp(2N)\times SU(3)_c\times  U(1)_Y.
    \label{eq:SU_breaking}
\end{equation}
There is a single $\bar\theta$ parameter in the UV for the gauge group $SU(2N+3)$ and thus only one axion is introduced to solve the strong CP problem since the underlying PQ symmetry, spontaneously broken at $f_a$, is assumed to be anomalous with respect to $SU(2N+3)$. The SM quarks, $q_i$ are embedded into fundamental representations of $SU(2N+3)$, $\Psi_i=(q_i,\psi_{q_i})^T$, where $\psi_{q_i}$ are exotic partners of the SM quarks and $i$ labels their flavor.  When $SU(2N+3)$ is broken down to $Sp(2N)\times SU(3)_c$, the exotic fermions $\psi_{q_i}$ transform as fundamentals of the $Sp(2N)$ gauge group, and the axion couples to both the $Sp(2N)$ and the $SU(3)_c$ gauge bosons. The Higgs sector is considered external to this breaking, with the electroweak and Yukawa interactions defined in terms of  $\Psi_i$. As such, the $\psi_{q_i}$ fermions have exactly the same Yukawa couplings as the SM quarks at the scale of the $SU(2N+3)$ breaking. We further assume that the $Sp(2N)$ gauge group confines at a scale, $\Lambda_{Sp}$ which is much larger than the electroweak scale.\footnote{If $SU(2N+3)$ is Higgsed down to $SU(2N)\times SU(3)_c$, the fermion condensates $\langle\psi_{q_i}\psi_{q_j}\rangle$ 
are charged under the electroweak gauge group and hence result in electroweak symmetry breaking at the scale $\Lambda_{Sp}\gg v_{SM}$. As such, the resulting gauge group must be $Sp(2N)$ or $SO(2N)$, because the fermion bilinear condensates can be electroweak singlets~\cite{DiVecchia:1980yfw, Bedi:2024kxe}.}
The axion, which is taken to be external to $SU(2N+3)$ dynamics, then mixes with the $Sp(2N)$ pions and obtains a mass.  For $f_a >\Lambda_{Sp}$, 
we can compute this contribution in the $Sp$ EFT  and is given by~\cite{Valenti:2022tsc}
\begin{equation}
    m_a^2f_a^2\simeq \frac{y_uy_d}{(16\pi^2)^2}\Lambda_{Sp}^4~,\label{eq:GC0}
\end{equation}
where $y_{u,d}$ are the Yukawa couplings for the up and down quark. The relation \eqref{eq:GC0} is shown by the dashed lines in \autoref{fig:summary} (bottom right). In contrast to the previous models,
in this model the hypercharge gauge group appears as a combination of $U(1)_{Y^\prime}$, and hence the axion necessarily couples to electroweak gauge bosons, implying  $c_{1,2} \sim c_3$.

An interesting modification of the model arises if we assume that the up-quark Yukawa coupling vanishes above the scale $\Lambda_{Sp}$. The confining dynamics of the $Sp$ gauge group can then explain the observed SM up quark mass, when additional vector-like fermions $\Psi_{{V}}$ (with the same gauge charges as $\Psi_{\bar{u}}$, containing the right-handed SM up quark, $\bar{u}$) are added, via spontaneous breaking of the chiral symmetry~\cite{Bedi:2024kxe}. Alternatively, we may assume the up-quark mass is generated by the SM up-quark Yukawa coupling, while the additional vector-like quark is massless. In this case, we do not need to introduce an elementary axion to the model by hand. Instead, one of the pions of the $Sp$ gauge sector can serve as an axion, with $f_a\simeq \Lambda_{Sp}/4\pi$. In this case the mass is given by 
\begin{equation}
    m_a^2\simeq  {y_uy_d}\frac{\Lambda_{Sp}^2}{16\pi^2}\simeq  {y_uy_d}\,f_a^2~.\label{eq:GCmass1}
\end{equation} This is shown by the solid red line in \autoref{fig:summary} (bottom right). The relation \eqref{eq:GCmass1} holds even for massive $\Psi_{ V}$, assuming $m_V < y_uy_d \Lambda_{Sp}/16\pi^2$,
where $m_V$ is the mass of $\Psi_{ V}$. However, for $f_a\sim ~$TeV and $m_V\sim~$TeV, we necessarily have  $m_V > y_uy_d \Lambda_{Sp}/16\pi^2\sim 10^{-11}f_a$. In this parameter space, we instead have 
\begin{equation}
    m^2_a\simeq m_V \Lambda_{Sp}\simeq 4\pi\,m_Vf_a~,\label{eq:GCmass2}
\end{equation}
similar to the Gell-Mann-Okubo mass formula in QCD. 
This mass relation is shown by the dotted lines in \autoref{fig:summary} (bottom right).
This composite axion couples to the gluons at tree level, owing to the PQ charges of the vector-like quarks, implying $c_3\sim 1.$ 
The $\eta^\prime$ of the $Sp$ sector couples to the electroweak gauge bosons due to the chiral anomaly and mixes with the axion, introducing couplings of the axion to the electroweak gauge bosons. This mixing is parametrically given by $m_a^2/\Lambda_{Sp}^2\sim m_V/(4\pi f_a)$, implying $c_{1,2}\sim 1$ for vector-like quark masses of approximately a few TeV and $f_a\sim 100~$GeV.

\begin{figure}[h]
    \centering
\includegraphics[width=0.9\linewidth]{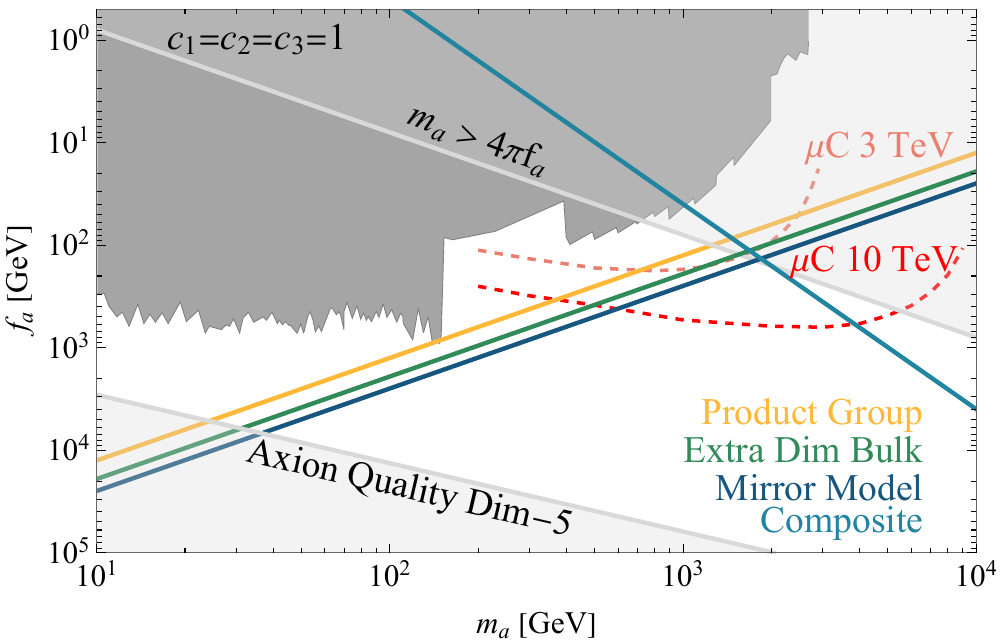}
    \caption{The projected sensitivity of a heavy QCD axion at a 3 and 10 TeV MuC shown as dashed lines. The dark-gray shaded region depicts the existing experimental constraints. 
    In the light gray region, labeled $m_a > 4\pi f_a$, the axion EFT under consideration breaks down, since depending on the UV completion, additional axion production and decay channels open up.
    Above the other light gray region labeled `Axion Quality Dim-5', there is no axion quality problem even from a dimension five operator of the type $\Phi^5/\mpl$. Thus, the white region in between the two light gray regions is highly motivated where both the strong CP and the axion quality problem can be solved.
    The four solid colored lines describe the four model benchmarks as discussed in~\autoref{sec:axion_model}. All of them provide targets for a 3 and 10 TeV MuC.
    }
    \label{fig:final_plot}
\end{figure}

\section{Conclusion}
\label{sec:conclusions}

In this work, we have studied the discovery potential of heavy QCD axions at a future high-energy muon collider. Heavy QCD axions are an interesting experimental target because they solve the strong CP problem and can also ameliorate the axion quality problem. In particular, a 10 TeV muon collider with 10~ab$^{-1}$ integrated luminosity can serve as a powerful probe of heavy QCD axions. The VBF processes dominate the axion production as expected, but both $Za$ associated and VBS channels yield non-negligible contributions, primarily due to their relatively lower backgrounds compared to the VBF signal search. Due to the axion-gluon coupling, the main axion decay channel is to gluons which gives rise to a di-jet final state.

Our phenomenological analysis focused on several different classes of motivated models that modify QCD dynamics at UV scales. 
This includes product group constructions~\cite{Agrawal:2017ksf, Gaillard:2018xgk,Csaki:2019vte}, extra-dimensional theories~\cite{Dienes2000,Gherghetta:2020keg}, mirror world scenarios~\cite{Rubakov:1997vp,Berezhiani:2000gh,Hook:2014cda,Fukuda:2015ana,Hook:2019qoh}, and color unification setups~\cite{Gherghetta:2016fhp, Valenti:2022tsc,Bedi:2024kxe}, all of which provide consistent mechanisms to enhance the axion mass while still preserving the solution to the strong CP problem. 
Our results, based on four model benchmarks, are shown in \autoref{fig:final_plot} for a 3 TeV and 10 TeV MuC. Note that the contribution from $Za$ production, which has a negligible impact on the projected sensitivity as discussed in \autoref{sec:production}, is not included in this plot. This is because in this channel $\sqrt{\hat{s}}=\sqrt{s}$ exceeds $4\pi f_a$ and new degrees of freedom could emerge in a UV model that leads to parameter-dependent interpretation of the projection.
For the ``Product-Group" scenario, we assume $M=5\times 10^{12}$~GeV, $\alpha_1(M)=0.12$, and $f_{a,2}=f_{a,3}=M$, similar to~\autoref{fig:ax_mass_SU(3)3}. 
The scenario where the axion arises from a bulk gauge field does not lead to a viable parameter space at a MuC. Instead, we model the axion as a bulk scalar to derive the ``Extra Dim Bulk" line, where we choose $1/R=2.5\times 10^6$~GeV and $\epsilon=0.36$, with two vector-like fermions as in~\eqref{eq:5d_scalar_mass2}. 
For the ``Mirror Model" line, we choose $\Lambda_{{\rm QCD}'}=500$~GeV. Finally, for the `Composite' line, we use~\eqref{eq:GCmass2} with $m_V=2$~TeV. There are two light gray regions, where the region labeled ``$m_a >4\pi f_a$", shows where the axion EFT breaks down 
while the region labeled ``Axion Quality Dim-5", depicts where there is an axion quality problem from dimension-5 operators such as $\Phi^5/\mpl$.
Note that while the axion EFT considered in this paper is not directly applicable in the upper gray region, the axion can still be studied in this region by assuming a UV completion for the axion-gluon coupling. This could lead to additional experimental signatures such as new axion production and decay channels, as well as, direct production of the radial mode associated with the complex scalar field containing the axion and heavy, exotic PQ-charged fermions.
Nevertheless, as can be seen in \autoref{fig:final_plot}, a muon collider can probe a significant region of parameter space (above the dashed lines) that significantly overlaps with each class of UV model (solid colored lines).
The sensitivity projections reveal that muon colliders can explore heavy axions in the multi-TeV regime, significantly extending the coverage compared to current experiments. This strongly motivates the muon collider as a discovery machine for QCD axions in scenarios that are both phenomenologically relevant and theoretically well-grounded.

Looking ahead, high-energy muon colliders provide a unique opportunity to explore heavy QCD axions. Their clean environment and high energy make them ideal for probing gluon-coupled axions that are otherwise difficult to access at hadron colliders. As collider designs and detector technologies advance, muon colliders could offer powerful sensitivity to heavy QCD axion models 
with $\lesssim\text{TeV}$ decay constants, opening a new window into the structure of QCD at high scales.

\acknowledgments

We thank Keisuke Harigaya, Kun-Feng Lyu, Lian-Tao Wang, and Andrea Wulzer for useful discussions. 
T.G., P.L., and Z.L., are supported by the DOE Grant No.~DE-SC0011842 at the University of Minnesota.
P.L. and Z.L. are also supported by a Sloan Research Fellowship from the Alfred P.~Sloan Foundation. R.B. is partly supported by Robert E. Greiling Jr. Fellowship at the University of Minnesota.
S.K. is supported in part by the NSF grant PHY-2210498 and the Simons Foundation. T.G. and Z.L. acknowledge the Munich Institute for Astro-, Particle and BioPhysics (MIAPbP), funded by the Deutsche Forschungsgemeinschaft (DFG, German Research Foundation) under Germany's Excellence Strategy – EXC-2094 – 390783311, where part of this work was done.

The data and core codes for the research are publicly available at \href{https://github.com/ZhenLiuPhys/axionMuC}{Github}.

\appendix
\section{Comparison Between Fixed-order Calculation and PDF Treatment}
\label{sec:calculation_VBS}

\begin{figure}[htbp]
    \centering
    \includegraphics[width=0.7\textwidth]{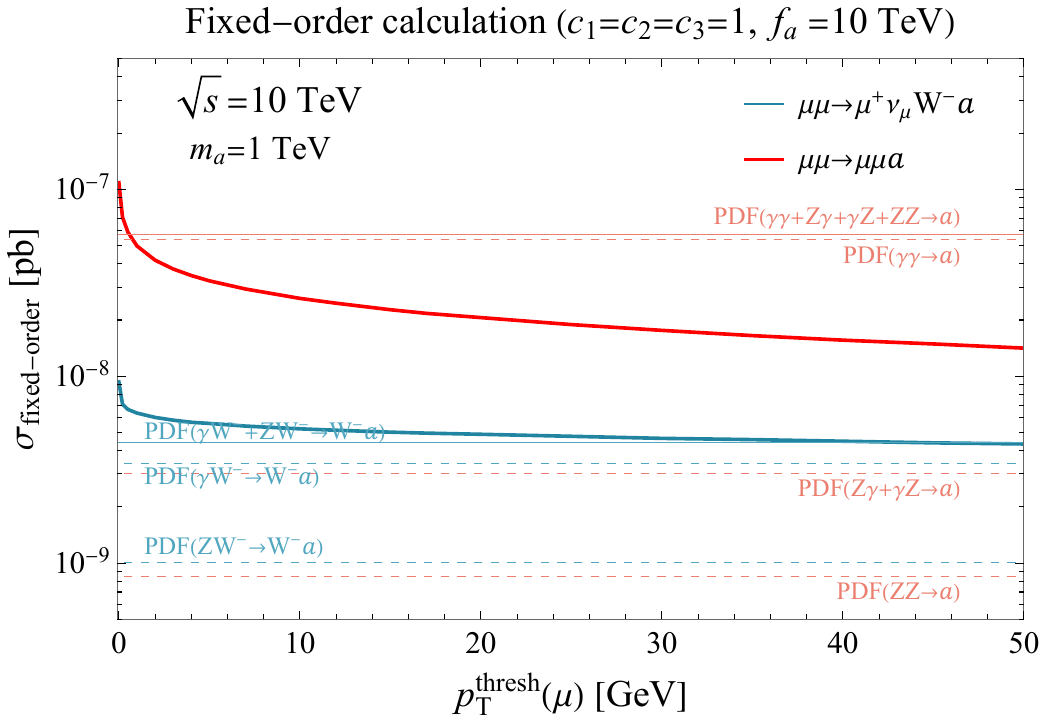}
    \caption{Fixed-order calculation of the VBS/VBF total cross section as a function of muon $p_T$ cut. 
    The thick-blue(red) curve is the fixed-order calculation of the 2-to-4(3) process as a function of the muon $p_T$ cut. The light-blue(red) horizontal line is the cross section of the 2-to-4(3) process via PDF (leading-log) treatment, which includes all partonic channels. Each dashed line corresponds to the result of a cross-section calculation involving the convolution with a single partonic channel. As expected, different processes have different IR divergences, which requires different IR regulators. For the 2-to-4 process, the fixed-order calculation matches with the PDF (leading-log) treatment around $p_T\sim40$~GeV. The VBF process has a slight divergence below the $\mathcal{O}(1)$~GeV threshold, and the matching scale between the PDF and the fixed-order calculation is roughly $\mathcal{O}(1)$~GeV $p_T$ cut. 
    }
    \label{fig:fixed_order_xs}
\end{figure}

\begin{figure}[htbp]
    \centering
    \includegraphics[width=0.7\textwidth]{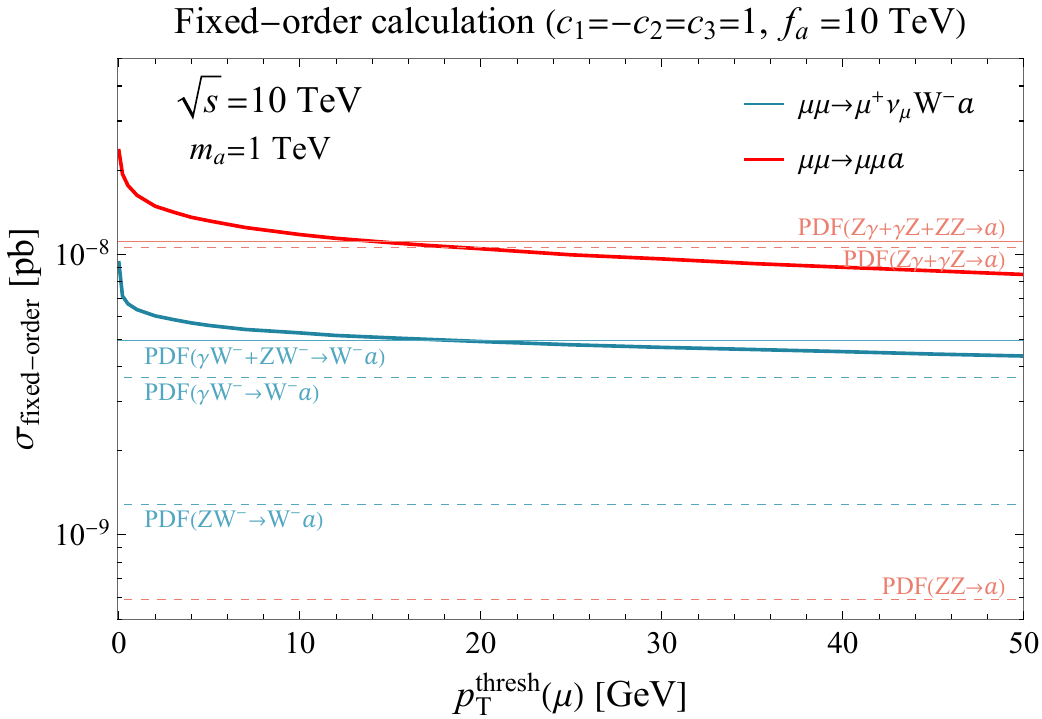}
    \caption{Fixed-order calculation of the VBS/VBF total cross section as a function of the muon $p_T$ cut. 
    The thick-blue(red) curve is the fixed-order calculation of the 2-to-4(3) process as a function of the muon $p_T$ cut. The light-blue(red) horizontal line is the cross section of the 2-to-4(3) process via PDF (leading-log) treatment, which includes all partonic channels. Each dashed line corresponds to the result of a cross-section calculation involving the convolution with a single partonic channel. As expected, different processes have different IR divergences, which requires different IR regulators. For the 2-to-4 process, the fixed-order calculation matches with the PDF (leading-log) treatment around $p_T\sim20$~GeV. The VBF process also has a matching scale around 15~GeV $p_T$ cut.}
    \label{fig:fixed_order_xs_nophoton}
\end{figure}

In this appendix, we select one representative VBF process and one representative VBS process to compare the fixed-order calculation with the PDF-based treatment in \autoref{fig:fixed_order_xs} and \autoref{fig:fixed_order_xs_nophoton}. The thick-solid curves in those figures depict the fixed-order calculation of the cross sections $\mu\mu\to\mu^+\nu_\mu W^- a$ and $\mu\mu\to \mu\mu a$ as a function of the muon $p_T$ cut. The cross sections calculated via the PDF treatment (including all partonic channels) are shown as light-solid lines, whereas each dashed line corresponds to the result of a cross section calculation involving the convolution with a single partonic channel.

In particular, \autoref{fig:fixed_order_xs} and \autoref{fig:fixed_order_xs_nophoton} use two benchmark coupling scenarios, one with and without axion-diphoton coupling, respectively. 
In \autoref{fig:fixed_order_xs}, the 2-to-3 process $\mu\mu\to\mu \mu a$ exhibits a ``matching scale'', where the fixed-order calculation matches the leading-log resummed PDF treatment, near a $p_T$ threshold of $\mathcal{O}(1)$~GeV. Below such a matching scale, the large logarithms from the fixed-order calculation need to be resummed, and above such a scale, higher-order fixed-order calculation with hard radiation above such a scale should be explicitly included. Note that in principle, the full evaluation of the fixed-order calculation should include $\mu\mu\rightarrow \mu\mu a +(n)\gamma$ to evaluate a more well-defined matching scale. Such a GeV matching scale is consistent with the photon PDF large log of $\ln(s/m_\mu^2)$. On the other hand, the 2-to-4 process $\mu\mu\to\mu^+ \nu_\mu W^- a$ reaches its matching scale near $\sim30~$GeV, which is compatible with the leading logs in such a process with $\ln(s/m_W^2)$. Note that $m_W^2$ is merely an approximate scale of the dominant momentum transfer $Q^2$ of these intermediate vector bosons. 
When choosing $c_1 = -c_2 = 1$ to eliminate the $\gamma\gamma$ fusion, as shown in \autoref{fig:fixed_order_xs_nophoton}, both processes exhibit a matching scale around $\sim 20$~GeV (see \autoref{fig:fixed_order_xs_nophoton}). These checks between fixed order and PDF treatment further validate the consistency of our treatment in this study.

\bibliographystyle{utphys}
\bibliography{references}

\end{document}